\documentclass[3p,times]{elsarticle}

\usepackage{amssymb}
\usepackage{amsmath}
\usepackage{mathtools}
\usepackage{graphicx}
\usepackage{bm}      
\usepackage{hyperref}
\usepackage{subfig}
\usepackage{makecell}
\usepackage{fontawesome}
\usepackage{tikz}
\usepackage{pgfplots}
\pgfplotsset{compat=1.16}

\usepackage{ulem}
\usepackage{cancel}
\usepackage{orcidlink}

\usepackage[figuresright]{rotating}

\newcommand*{\complexi}{\mathrm{i}}
\newcommand*{\transpose}{\mathbf{T}}
\newcommand{\mat}[1]{\bm{#1}}
\newcommand{\sapphire}{\texttt{Sapphire++}\ }
\newcommand{\sapphiren}{\texttt{Sapphire++}}

\renewcommand{\emph}{\textit}
\DeclareMathOperator\erfc{erfc}

\begin{document}

\begin{frontmatter}
  \title{Sapphire++: A particle transport code combining a spherical harmonic expansion and the discontinuous Galerkin method}

  \author[MPIK]{Nils W. Schween\orcidlink{0000-0003-0105-8733}\corref{cor1}}
  \ead{nils.schween@mpi-hd.mpg.de}
  \author[MPIK]{Florian Schulze\orcidlink{0000-0003-1644-8697}}
  \author[MPIK]{Brian Reville\orcidlink{0000-0002-3778-1432}}

  \affiliation[MPIK]{organization={Max-Planck-Institut für Kernphysik},
    addressline={Saupfercheckweg 1},
    postcode={69117},
    city={Heidelberg},
    country={Germany}}

  \cortext[cor1]{Corresponding author}
  \begin{abstract}
    We present \sapphiren, an open-source code designed to numerically solve the Vlasov--Fokker--Planck equation for astrophysical applications. \sapphire employs a numerical algorithm based on a spherical harmonic expansion of the distribution function, expressing the Vlasov--Fokker--Planck equation as a system of partial differential equations governing the evolution of the expansion coefficients. The code utilises the discontinuous Galerkin method in conjunction with implicit and explicit time stepping methods to compute these coefficients, providing significant flexibility in its choice of spatial and temporal accuracy. We showcase the code's validity using examples. In particular, we simulate the acceleration of test particles at a parallel shock and compare the results to analytical predictions. The \sapphire code {\href{https://github.com/sapphirepp/sapphirepp}{\faGithub}} is available as a free and open-source tool for the community.
  \end{abstract}


\end{frontmatter}

\section{Introduction}
\label{sec:introduction}

In laboratory and astrophysical settings it is frequently necessary to calculate the transport of charged particles or photons in an inhomogeneous medium or plasma in which they are scattered. Most realistic scenarios are modelled with equations that cannot be solved analytically, and a numerical solution is required. To this end, we have developed a new, free and open-source code, \sapphire (``\textbf{S}imulating \textbf{a}strophysical \textbf{p}lasmas and \textbf{p}articles with \textbf{hi}ghly \textbf{r}elativistic \textbf{e}nergies in \texttt{C\textbf{++}}'')\footnote{\url{https://sapphirepp.org}}. In this paper we detail the numerical algorithms in \sapphire and explore the capabilities of the code by means of four physically motivated examples. As the acronym \sapphire implies, the code is written in the \texttt{C++} programming language and is developed for simulating highly energetic charged particles that interact with astrophysical plasmas. The approach can however be applied also to non-relativistic particles, and the algorithms can in principle be modified to simulate photon/neutrino transport.

In \sapphire the propagation and acceleration of charged particles in a prescribed background plasma is modelled with a Vlasov--Fokker--Planck (VFP) equation, formulated in a mixed-coordinate system, i.e. the momenta of the particles are defined in the rest frame of the plasma which moves with velocity $\mathbf{U}$ in a fixed laboratory frame. This is done to simplify the collision operator, which we assume to model elastic, isotropic scattering in the local fluid frame, i.e. the rest frame of the background plasma. The VFP equation in this mixed-coordinate system is, to first order in $U/c$ (see for example \cite{Achterberg2018_1})
\begin{equation}
  \label{eq:vfp-mixed-coordinates}
  \left(1+\frac{\mathbf{U}\cdot \mathbf{V}'}{c^2}\right)\frac{\partial f}{\partial t}
  + \left(\mathbf{U} + \mathbf{V}'\right) \cdot \nabla_{x} f
  - \left(\gamma' m \frac{\mathrm{d} \mathbf{U}}{\mathrm{d} t}
  + (\mathbf{p}' \cdot \nabla_{x}) \mathbf{U} \right) \cdot \nabla_{p'} f
  + q \mathbf{V}' \cdot \left(\mathbf{B}' \times \nabla_{p'} f \right)
  = \frac{\nu'}{2}\Delta_{\theta', \varphi'}f \,,
\end{equation}
where the primed quantities are given in the rest frame of the background (magnetized) fluid.

\sapphire computes the single particle distribution function $f$ that describes the phase space density of an energetic particle species with rest-mass $m$ and charge $q$. It is assumed that this species is distinct from the background plasma that supports the electromagnetic fields which mediate the scattering. In eq. \eqref{eq:vfp-mixed-coordinates},  $\mathbf{V}'$ represent the particle velocity, $\mathbf{p}'=\gamma'm \mathbf{V}'$ its momentum, and $\mathbf{B}'$ is the mean magnetic field threading the background plasma through which the energetic particles propagate. We assume for the remainder of this paper that the ideal magnetohydrodynamic (MHD) approximation applies, and thus the electric field vanishes in the local fluid frame, i.e. $\mathbf{E}'= -\mathbf{U}' \times \mathbf{B}' = 0$. As noted in \cite[Sec. 2]{Schween24a} this restriction can be relaxed, and alternative Ohm's laws can be implemented. We note that the velocity field $\mathbf{U}$ and the magnetic field $\mathbf{B}'$ of the background plasma are currently prescribed by the user in \sapphire, i.e. the test particle limit is assumed. This means that there is no energy exchange between the accelerated particles and the background plasma. Future versions will compute these fields self-consistently by co-evolving the equations of MHD \cite[e.g.][]{Reville2013}. The right-hand side of equation \eqref{eq:vfp-mixed-coordinates} models the change of the distribution function $f$ due to interactions of the particles with stochastic electromagnetic fluctuations (MHD turbulence) in the background plasma, i.e. due to being scattered while traversing the plasma. The presented form of the collision operator implies that we only consider collisions that conserve energy in the local fluid frame, i.e. elastic scattering. Its rate is set by the scattering frequency $\nu'(p')$, which, for simplicity, we take to be only a function of the momentum's magnitude. The angular part of the Laplacian operator implies ``diffusion'' in the angular variables $\theta'$ and $\varphi'$ of $\mathbf{p'} = p' \left(\cos\theta', \sin\theta' \cos\varphi', \sin\theta' \sin\varphi' \right)^{T}$. This diffusion is interpreted as a result of many small changes in the particles' direction of motion. More details about the mixed-coordinate system can be found in~\cite{Achterberg2018_1} and references therein.

For the rest of the paper we drop the primes appearing in equation~\eqref{eq:vfp-mixed-coordinates}, and it is henceforth understood that $V, \gamma, p, \theta, \varphi$ and $\nu$ are referring to quantities defined in the rest frame of the background plasma. Moreover, we also drop the relativistic correction to the time derivative of the single particle distribution function, i.e. $\mathbf{U}\cdot \mathbf{V}'/c^2$. We note that its inclusion is necessary to ensure accuracy to first order in $U/c$ for \emph{time-dependent} problems (see \ref{app:high-order}).

In order to reduce the dimensionality of the problem, \sapphire treats the momentum phase space using a truncated spherical harmonic expansion i.e.
\begin{equation}
  \label{eq:spherical-harmonic-exp}
  f(t, \mathbf{x}, p, \theta, \varphi) =
  \sum^{l_{\text{max}}}_{l = 0} \sum^{l}_{m = 0 } \sum^{1}_{s = 0} f_{lms}(\mathbf{x}, p, t) Y_{lms}(\theta, \varphi) \, ,
\end{equation}
where the $Y_{lms}$ are the real spherical harmonics (see \ref{app:definition-real-spherical-harmonics} for a definition) and $l_{\text{max}}$ is the maximum order of the expansion. Computing $f$ now amounts to computing the expansion coefficients $f_{lms}$. Substituting eq.~\eqref{eq:spherical-harmonic-exp} into eq.~\eqref{eq:vfp-mixed-coordinates}, one can derive a system of partial differential equations (PDEs) for the $f_{lms}$. The result of this derivation is summarised in eq.~\eqref{eq:real-system-of-equations} of Section~\ref{sec:system-of-equations}, and for further details on the derivation, we refer the reader to our companion paper \cite{Schween24a}, \cite[cf.][]{Bell2011, Reville2013}.

If the particles are scattered frequently enough to ensure that the characteristic length-scale $\mathcal{L}$ for spatial gradients of the isotropic part $f_{000}$ of the distribution function is long relative to the scattering mean free path $\lambda = V/\nu$ of the particles, the diffusion approximation is used. This means that the spherical harmonic expansion is truncated at $l_{\text{max}} = 1$ and that $f_{100}, f_{110}$ and $f_{111}$ are assumed to be of the order $\mathcal{O}(\lambda/\mathcal{L} f_{000})$. In such situations, the system of PDEs can be reduced to a simple advection-diffusion equation for $f_{000}$. In the astrophysics literature this equation is commonly called the cosmic-ray transport equation \cite{Ginzburg, Parker, Skilling1975, Webb1989, Kirk88}. However, in many physical situations one encounters circumstances in which the diffusion approximation is not applicable, i.e. it is necessary to include more terms of the spherical harmonic expansion and drop the assumption that these terms are small in comparison to the isotropic part. The choice of a large $l_{\text{max}}$ in \sapphire allows for such a treatment.

Numerical solutions to the VFP equation, coupled with Maxwell's equations are frequently used in laboratory plasma studies where Coulomb collisions play an important role. This is particularly important for inertial confinement fusion investigations; see for example the KALOS \cite{Bell2006}, IMPACT \cite{IMPACT} and OSHUN \cite{Tzoufras2011} codes; also \cite{Zhang24,Bell24} for more recent developments. These codes typically apply either finite difference \cite{Bell2006} or finite volume approaches \cite{Zhang24}. In contrast, \sapphire solves the system of equations with the discontinuous Galerkin (dG) method. The dG method is a finite element (FE) method, ideally suited for advection-reaction equations. The dG method is also used in the \texttt{Gkeyll} code \cite{Hakim2020} to solve the Vlasov equation in Cartesian spatial and momentum coordinates using (up to) 6D discontinuous basis functions. \sapphire tries to combine the advantages of both, i.e. the expansion of the distribution function and the dG method. The implementation of \sapphire is based on the FE library \texttt{deal.ii}\footnote{\url{https://www.dealii.org/}} \cite{dealii2019, dealii95}.

In section \ref{sec:system-of-equations} we apply the operator based method developed previously in \cite{Schween24a} to the VFP equation \eqref{eq:vfp-mixed-coordinates} to arrive at a system of PDEs for the expansion coefficients. The process of discretising this system applying the dG method is described in section \ref{sec:discontinuous-galerkin}. In section \ref{sec:tests} we present simulations of several test cases, including the acceleration of particles at a parallel shock, and compare the results with known analytic solution from diffusive shock acceleration theory. Our conclusions are presented in section \ref{sec:conclusions}.

\section{The system of partial differential equations}
\label{sec:system-of-equations}

In this section we present the system of PDEs for the expansion coefficients $f_{lms}$. The details of its derivation can be found in \cite{Schween24a}. The system of PDEs we solve in \sapphire can be expressed as
\begin{equation}
  \label{eq:real-system-of-equations}
  \begin{split}
    \partial_{t}\mathbf{f}
     & + \left(U^{a}\mat{1} + V \mat{A}^{a}\right) \partial_{x^{a}}\mathbf{f}
    - \left(\gamma m \frac{\mathrm{d} U_{a}}{\mathrm{d} t} \mat{A}^{a}
    + p \frac{\partial U_{b}}{\partial x^{a}} \mat{A}^{a}\mat{A}^{b}\right)\partial_{p} \mathbf{f}             \\
     & {} + \left(\frac{1}{V} \epsilon_{abc} \frac{\mathrm{d} U^{a}}{\mathrm{d} t} \mat{A}^{b}\mat{\Omega}^{c}
    + \epsilon_{bcd} \frac{\partial U_{b}}{\partial x^{a}}\mat{A}^{a}\mat{A}^{c}\mat{\Omega}^{d}\right) \mathbf{f}
    - \omega_{a}\mat{\Omega}^{a}\mathbf{f}
    + \nu \mat{C}\mat{f} = 0 \,,
  \end{split}
\end{equation}
where we introduced the relativistic gyro-frequency vector $\omega_a = qB_a/\gamma m$. Summation over repeated indices in eq.~\eqref{eq:real-system-of-equations} is implied. As stated in the introduction, we have dropped the relativistic correction to the time derivative in moving from eq.~\eqref{eq:vfp-mixed-coordinates} to eq.~\eqref{eq:real-system-of-equations}, and as such, any solutions which are time dependent are accurate only to zeroth order in $U/c$, i.e. corrections of order $U/c$ are dropped (see \cite{Skilling1975} for discussion). We discuss approaches to recover higher order accuracy in \ref{app:high-order}.

The remaining terms in eq.~\eqref{eq:real-system-of-equations} including the vector $\mathbf{f}$, which contains the expansion coefficients, and the matrices $\mat{A}^{a}, \mat{\Omega}^{a}$ and $\mat{C}$ are defined in the next section.

\subsection{Explicit expression for the system matrices}
\label{sec:definitions-system-of-equations}

The symbols $\mat{A}^{a}, \mat{\Omega}^{a}$ and $\mat{C}$ denote \emph{real} matrices that are elements of $\mathbb{R}^{n \times n}$ with $n = \sum^{l_{\text{max}}}_{l=0} (2l + 1) = (l_{\text{max}} + 1)^{2}$ and $l_{\text{max}}$ is the degree at which the spherical harmonic expansion~\eqref{eq:spherical-harmonic-exp} is truncated. The vector $\mathbf{f}$ has components
\begin{equation}
  \label{eq:vector-expansion-coefficients}
  \left(\, \mathbf{f}\right)_{j(l,m,s)} \coloneqq f_{lms}  \,,
\end{equation}
where $j(l,m,s)$ is a one-to-one function of the indices $l, m$ and $s$. We call this function, which determines how the expansion coefficients are ordered, an \textit{index map}. In \sapphire we choose the following ordering
\begin{equation}
  \label{eq:ordering-expansion-coefficients}
  \mathbf{f} = (f_{000}, f_{110}, f_{100}, f_{111}, f_{220}, f_{210}, f_{200}, f_{211}, f_{221} \dots)^{\transpose}\,.
\end{equation}
The corresponding index map is $j(l, m, s) = l(l + 1) + (-1)^{s + 1} m$ with $j$ starting at zero.

The matrix elements of $\mat{\Omega}^{x}$ and $\mat{A}^{x}$ are
\begin{align}
  \label{eq:representation-matrix-Omegax}
  (\mat{\Omega}^{x})_{i(l',m',s')j(l,m,s)} & = m \delta_{l'l}\delta_{m'm}
  \left(\frac{\delta_{s'0}\delta_{s1}}{\sqrt{1 + \delta_{m'0}}} - \frac{\delta_{s'1}\delta_{s0}}{\sqrt{1 + \delta_{m0}}}\right) \quad \text{and} \\
  \label{eq:representation-matrix-Ax}
  (\mat{A}^{x})_{i(l',m',s')j(l,m,s)}      & = \delta_{m'm}\delta_{s's}
  \left(\sqrt{\frac{(l + m + 1)(l - m +1)}{(2l + 3)(2l + 1)}} \delta_{l'(l+1)} + \sqrt{\frac{(l+m)(l-m)}{(2l + 1)(2l - 1)}} \delta_{l'(l-1)}\right) \,.
\end{align}
The matrices $\mat{A}^{y}, \mat{A}^{z}, \mat{\Omega}^{y}$ and $\mat{\Omega}^{z}$ can be related to one another through rotation matrices, i.e.
\begin{alignat}{2}
  \label{eq:real-direction-op-matrices-and-rotations}
  \mat{A}^{y}      & =
  \mathrm{e}^{-\frac{\pi}{2}\mat{\Omega}^{z}}\mat{A}^{x}\mathrm{e}^{\frac{\pi}{2}\mat{\Omega}^{z}}
  \qquad           &
  \mat{A}^{z}      & =
  \mathrm{e}^{-\frac{\pi}{2}\mat{\Omega}^{x}}\mat{A}^{y}\mathrm{e}^{\frac{\pi}{2}\mat{\Omega}^{x}} \quad \text{and}                \\
  \label{eq:real-angular-momentum-op-matrices-and-rotations}
  \mat{\Omega}^{y} & = \mathrm{e}^{-\frac{\pi}{2}\mat{\Omega}^{z}}\mat{\Omega}^{x}\mathrm{e}^{\frac{\pi}{2}\mat{\Omega}^{z}}     &
  \mat{\Omega}^{z} & = \mathrm{e}^{-\frac{\pi}{2}\mat{\Omega}^{x}}\mat{\Omega}^{y}\mathrm{e}^{\frac{\pi}{2}\mat{\Omega}^{x}} \,.
\end{alignat}
We note that the $\mat{A}$ matrices are symmetric whereas the $\mat{\Omega}$ matrices are antisymmetric.

The matrix elements of the rotation matrices are
\begin{equation}
  \label{eq:representation-matrix-Ux}
  \begin{split}
    (\mathrm{e}^{\frac{\pi}{2}\mat{\Omega}^{x}})_{i(l',m',s')j(l,m,s)}
    = \delta_{l'l}\delta_{m'm} & \left[\frac{\delta_{s'0}\delta_{s0}}{\sqrt{(1 + \delta_{m'0}) (1 + \delta_{m0})}} \left(\cos\left(\frac{\pi}{2}m\right)  + 1 \delta_{m'0}\delta_{m0}\right) \right. \\
                               & \left. {} + \frac{\delta_{s'1}\delta_{s0}}{\sqrt{1 + \delta_{m'0}}} \sin\left(\frac{\pi}{2} m\right)
    - \frac{\delta_{s'0}\delta_{s1}}{\sqrt{1 + \delta_{m0}}} \sin\left(\frac{\pi}{2} m\right)
    + \delta_{s'1}\delta_{s1} \cos\left(\frac{\pi}{2} m\right)\right]
  \end{split}
\end{equation}
and
\begin{equation}
  \label{eq:representation-matrix-Uz}
  \begin{split}
    (\mathrm{e}^{\frac{\pi}{2}\mat{\Omega}^{z}})_{i(l',m',s')j(l,m,s)}
     & =  \frac{\delta_{s'0}\delta_{s0}}{\sqrt{(1 + \delta_{m'0}) (1 + \delta_{m0})}}\left[u(l',m', l, m) + (-1)^{m}u(l',m', l, -m) \right] \\
     & \phantom{=} {}+ \delta_{s'1}\delta_{s1}\left[u(l',m', l, m) - (-1)^{m} u(l',m', l, -m) \right] \,,
  \end{split}
\end{equation}
where we introduced the function
\begin{equation}
  \label{eq:complex-rotation}
  u(l',m', l,m) \coloneqq \delta_{l'l}  \frac{(-1)^{l - m'}}{2^{l}} \sum^{n}_{\substack{k = 0\\ m' + m + k \geq 0}} (-1)^{k}
  \frac{\left[(l + m')!(l - m')!(l + m)!(l - m)!\right]^{1/2}}{k!(l - m' - k)!(l - m - k)!(m' + m + k)!} \,.
\end{equation}
Due to the many factorials, care is required for a stable implementation of the rotation matrices; see \cite{Choi1999} for an example implementation. In \sapphire we use explicit expressions for all matrices, which we documented previously \cite[Appendix B]{Schween24a}.

It is left to give an expression for the \textit{collision matrix} $\mat{C}$, which is a diagonal matrix whose elements are
\begin{equation}
  \label{eq:real-C-matrix}
  (\mat{C})_{i(l'm's')j(l,m,s)} = \frac{l(l + 1)}{2}\delta_{l'l}\delta_{m'm}\delta_{s's} \,.
\end{equation}

As discussed in \cite[Sec. 3.4]{Schween24a}, the derivation of the system of PDEs~\eqref{eq:real-system-of-equations} requires that the truncation of the spherical harmonic expansion~\eqref{eq:spherical-harmonic-exp} be made after determining the matrix elements. If it is truncated at $l_{\text{max}}$ at the outset, the evaluation of the matrix products, for example $\mat{A}^{a}\mat{A}^{b}$, requires that all involved matrices are constructed for $L = l_{\text{max}} + 1$ before multiplying them. The resulting matrix can then be reduced to a matrix corresponding to $l_{\text{max}}$ by extracting a submatrix whose size is $n \times n$. Furthermore, the submatrix corresponding to $\mat{A}^{a}\mat{A}^{b}$ is symmetric.

\subsection{Advection-reaction equation, boundary conditions and initial conditions}
\label{sec:advection-reaction-equation-boundary-conditions}

In the last part of this section, we express the system of equations~\eqref{eq:real-system-of-equations} as an \textit{advection-reaction equation}. The advection-reaction equation is known to be well suited for an application of the dG method, \cite{Pietro_MathematicalAspectsDG}. The system of PDEs can be brought into the following form
\begin{equation}
  \label{eq:system-as-advection-reaction-eq}
  \partial_{t}\mathbf{f} + (\mat{\beta} \cdot \tilde{\nabla}) \mathbf{f}
  + \mat{R} \mathbf{f} = 0 \,,
\end{equation}
where we introduce the symbols
\begin{equation}
  \label{eq:definitions-advection-reaction-eq}
  \begin{split}
    (\mat{\beta})^{\alpha}    & \coloneqq
    \begin{cases}
      U^{\alpha}\mat{1} + V \mat{A}^{\alpha}                           & \text{for } \alpha \in \{1, 2, 3\} \\
      - \gamma m \frac{\mathrm{d} U_{a}}{\mathrm{d} t} \mat{A}^{a}
      - p \frac{\partial U_{b}}{\partial x^{a}} \mat{A}^{a}\mat{A}^{b} & \text{for } \alpha = 4
    \end{cases} \,, \\
    (\tilde{\nabla})_{\alpha} & \coloneqq
    \begin{cases}
      \partial/\partial x^{\alpha} & \hspace{7.51em} \text{for } \alpha \in \{1, 2, 3\} \\
      \partial /\partial p         & \hspace{7.51em} \text{for } \alpha = 4
    \end{cases}\text{ and}                                                                       \\
    \mat{R}                   & \coloneqq
    \frac{1}{V} \epsilon_{abc} \frac{\mathrm{d} U^{a}}{\mathrm{d} t} \mat{A}^{b}\mat{\Omega}^{c}
    + \epsilon_{bcd} \frac{\partial U_{b}}{\partial x^{a}}\mat{A}^{a}\mat{A}^{c}\mat{\Omega}^{d}
    - \omega_{a}\mat{\Omega}^{a}
    + \nu \mat{C} \,.
  \end{split}
\end{equation}
We call the $\mat{\beta}^{\alpha}$ the \textit{advection matrices} and $\mat{R}$ is the \textit{reaction matrix}. Moreover, we refer to the vector space $\bm{\xi}=(\mathbf{x}, p)^{\transpose}$ as \textit{reduced phase-space}, and use the index $\alpha$ to refer to the components of vectors in the reduced phase-space. The above system of partial differential equations is a \textit{linear hyperbolic system}, because the advection matrices are symmetric, which implies that they are diagonalisable and that their eigenvalues are real, see, for example, \cite[Sec. 2.9]{LeVeque_FVM}. Note that due to the symmetry of $\mat{A}^{a}$ and of the products $\mat{A}^{a}\mat{A}^{b}$, the advection matrices $\mat{\beta}^{\alpha}$ are also symmetric.

The physical interpretation of the terms in the advection-reaction equations is, firstly, that a combination of expansion coefficients $\mathbf{f}$ is advected in the reduced phase-space variables $\bm{\xi}$ and, secondly, the expansion coefficients are mixed and decay through the reaction matrix $\mat{R}$.

A unique solution of the system of PDEs~\eqref{eq:system-as-advection-reaction-eq} requires in addition boundary and initial conditions. In the subsequent section we choose a zero inflow boundary condition, because the mathematical results concerning the uniqueness of the solution, to which we refer the reader, hold for this choice. However, in \sapphire other boundary conditions are implemented, namely periodic and continuous, which we use in the examples in section~\eqref{sec:tests}. We denote with $\mathbf{f}^{-}$ the inflowing part of $\mathbf{f}$ at a specific point on the boundary $\partial D$ of the domain $D \subset \mathbb{R}^{4}$. Zero inflow at all times is then formally expressed as $\mathbf{f}^{-} = 0$ on $ \partial D \times [0, t_{F}]$ where $t_{F}$ is the final time. We address the question of how to determine $\mathbf{f}^{-}$ later, see eq.~\eqref{eq:zero-inflow-boundary} and the explanations thereafter. As an initial condition we choose a smooth function $\mathbf{f}(\mathbf{x}, p,t=0)=\mathbf{f}^{0}(\mathbf{x}, p)$.

Furthermore, we include an additional source term $\mathbf{s}(\mathbf{x}, p, t)$, which is at least a square-integrable function. The problem we seek to solve thus becomes
\begin{alignat}{2}
  \partial_{t}\mathbf{f} + (\mat{\beta} \cdot \tilde{\nabla})\mathbf{f} + \mat{R} \mathbf{f} & = \mathbf{s}\qquad &  & \text{in } D \times [0, t_{F}]    \label{eq:well-posed-advection-reaction-eq} \\
  \mathbf{f}^{-}                                                                             & = 0                &  & \text{on }  \partial D \times [0, t_{F}]\label{eq:zero-inflow-bc}             \\
  f(\mathbf{x}, p, 0)                                                                        & = \mathbf{f}^{0}   &  & \text{in } D\label{eq:initial-condition} \,.
\end{alignat}
Under the assumption that $\bm{\beta}$ and $\mat{R}$ do not depend on time it can be shown via an \textit{energy estimate}, that if a solution $\mathbf{f}$ exists, the solution is unique, see \cite[p.70 Lemma 3.2 and p.332 Lemma 7.26]{Pietro_MathematicalAspectsDG}.\footnote{ We note that Di Pietro \& Ern would interpret such a system as an example of a \textit{Friedrich's system}, cf. \cite[Section 7.1 and Section 7.5]{Pietro_MathematicalAspectsDG}}

\section{Discontinuous Galerkin}
\label{sec:discontinuous-galerkin}

As mentioned, the advection-reaction system is (in the test-particle limit under consideration) a linear hyperbolic system. A well established approach to solve such a system numerically is the finite volume (FV) method, because it includes fluxes which allow it to conserve the relevant physical quantities. We apply the discontinuous Galerkin method instead of the FV method, because the dG method is also based on fluxes and, thus, has the same main advantage but, in contrast to FV methods, it is easy to increase the order of accuracy of the spatial discretisation of the solution to the PDE system~\eqref{eq:system-as-advection-reaction-eq}. FV methods rely on polynomial reconstruction methods like the (weighted) essentially non-oscillatory (WENO) method, which requires a stencil of cells to achieve higher order accuracy ~\cite[e.g.][]{Shu2009}. As we show in this section, the dG method works right away with higher order polynomials, which are independently defined on each cell, thus avoiding the need for a reconstruction algorithm with information from neighbouring cells. In this sense the dG method is more \textit{local} than FV methods. It is this locality which helps to leverage the implementation of algorithms which adapt the cell sizes or the polynomial degree depending on the error of the numerical solution. This is useful in the context of the acceleration of particles around a shock, which benefits from high accuracy in the vicinity of the shock, see for example the grid design in Fig.~\ref{fig:shock-grid}. Future versions will exploit this advantage more completely and employ the facilities of the \texttt{deal.ii} library to implement adaptive mesh refinement.

We next explain how to apply the dG method to the system of PDEs~\eqref{eq:system-as-advection-reaction-eq} and how we can exploit properties of the $\mat{A}^{a}$ matrices to accelerate and stabilise its solution. The aim of the explanations is twofold: We would like to provide \sapphire users with a detailed description of the spatial discretisation algorithm and to present the material in a way that is accessible to physicists and applied mathematicians. We note that the content up to eq.~\eqref{eq:definition-upwind-flux} heavily draws from \cite[in particular Chap. 1 -- 3 and Chap. 7]{Pietro_MathematicalAspectsDG}. Readers familiar with the dG method can directly jump to the definition of the numerical flux in eq.~\eqref{eq:definition-upwind-flux}, where we introduce a novel way to compute the upwind flux at the cell interfaces.

\subsection{Discrete representation of the solution and the finite element method}
\label{sec:discrete-representation-solution}

The dG method is a finite element method (FEM) in which the discrete approximation of the solution to the PDE is represented by a linear combination of functions, i.e.
\begin{equation}
  \label{eq:dg-discrete-representation-solution}
  \mathbf{f}_{h}(t, \mathbf{x}, p) =  \zeta_{j}(t) \bm{\phi}_{j}(\mathbf{x}, p)
  \quad \text{with } \bm{\phi}_{k} \in V_{h} \,,
\end{equation}
where summation over $j$ is implied. $V_{h}$ is a finite dimensional function space and the $\bm{\phi}_{j}$ are its basis functions. The subscript $h$ refers to a typical cell size and expresses that $V_{h}$ depends among other things on the number of cells in which the domain $D$ is decomposed. The objective of FEMs is to determine the coefficients $\zeta_{j}$, which are called \textit{degrees of freedom} (DoF).

One of the distinguishing features of a dG method is the choice of the function space $V_{h}$: The domain $D$ is subjected to a triangulation. The outcome is a set of cells in the reduced phase-space which we denote with $\mathcal{T}_{h}$. In \sapphire these are lines in one dimension, rectangles in two dimensions and cuboids in three dimensions\footnote{Hypercuboids can be applied in higher dimensions, though \sapphire is not yet equipped to handle these within the standard \texttt{deal.ii} framework. Extension to higher dimensions are implemented in \texttt{hyper.deal}, as described in \cite{munch2020}.}\,. Subsequently, a set of functions is defined on each cell $T$, for example, polynomials up to a certain degree $k$. In \sapphire a \textit{tensor product} of 1D \textit{Lagrange polynomials} is used. For $k\leq 2$, the Lagrange polynomials are constructed with equidistant points. For $k >2$ \textit{Gauss--Lobatto} points are used; see the \texttt{deal.ii} manual~(\href{https://dealii.org/9.4.1/doxygen/deal.II/classFE__Q.html}{FE\_Q}).

The $k+1$ Gauss--Lobatto points are found by combining the roots of the derivative of the degree-$k$ Legendre polynomial $P'_{k}(x)$ with the interval endpoint $\{-1, 1\}$, see, for example,~\cite[eq. 25.4.32]{Stegun_HandbookOfFunctions} or~\cite[p. 47]{Hesthaven_NodalDG}. The corresponding Lagrange polynomial basis is
\begin{equation}
  \label{eq:lagrange-polynomial}
  \ell_{i}(x) = \prod_{\substack{0 \leq j \leq k \\ i \neq j}} \frac{x - x_{j}}{x_{i} - x_{j}}
  \quad\text{where the } x_{i} \text{ are the } k + 1
  \text{ Gauss--Lobatto points, see, for example,~\cite[eq. 25.2.2]{Stegun_HandbookOfFunctions}} \,.
\end{equation}

Since the expansion coefficients $\mathbf{f}$ of the spherical harmonic expansion depend on $\bm{\xi} \in \mathbb{R}^{d+1}$, a representation of them in terms of polynomials requires polynomials of $d+1$ variables. These polynomials are constructed by taking the tensor product of the Lagrange polynomial bases, which in this context is just the ordinary product of the polynomials. For example, for $d =2$ the set of functions defined on a cell is
\begin{equation}
  \label{eq:polynomial-space-on-one-cell}
  \mathbb{Q}^{k}(T) =\text{span}\{\ell_{i}(x)\ell_{j}(y)\ell_{m}(p) \}\qquad\text{with }
  i,j,m \in \{0, \dots, k\} \text{ and } \bm{\xi} \in T \subset \mathbb{R}^{d + 1} \,.
\end{equation}
We note that $\mathbb{Q}^{k}(T)$ is a vector space with dimension $\dim(\mathbb{Q}^{k}(T)) = (k + 1)^{d+1}$.

The space $\mathbb{Q}^{k}(T)$ can be used to represent \emph{one} of the $n = (l_{\text{max}} + 1)^{2}$ expansion coefficients $f_{lms}$. Hence, we require a copy of $\mathbb{Q}^{k}(T)$ for each expansion coefficient. This requirement is condensed in the introduction of the space $[\mathbb{Q}^{k}(T)]^{n}$ with dimension $\dim([\mathbb{Q}^{k}(T)]^{n}) = n (k + 1)^{d+1}$. $[\mathbb{Q}^{k}(T)]^{n} $ is a space with vectors $\mathbf{v} \in \mathbb{R}^{n}$ whose components are elements of $\mathbb{Q}^{k}(T)$, i.e. they are linear combinations of the products of the Lagrange polynomials.

Eventually, the space $V_{h}$ can be defined as the direct sum over of all cells $[\mathbb{Q}^{k}(T)]^{n}$, i.e.
\begin{equation}
  \label{eq:finite-element-space}
  V_{h} \coloneqq \bigoplus_{T \in \mathcal{T}_{h}} [\mathbb{Q}^{k}(T)]^{n} \,.
\end{equation}
This means that every element in $V_{h}$ is a sum of the polynomials defined on each cell. Since there is \emph{no} requirement that this sum has to give a continuous function at the cell interfaces, elements of $V_{h}$ are expected to be discontinuous at cell faces. Such a space is called a \textit{broken polynomial space}, see, for example, \cite[Sections 1.2.4.2 - 3]{Pietro_MathematicalAspectsDG}. Moreover, it is from this that the \emph{discontinuous} Galerkin name arises.

As stated in the definition of the discrete solution~\eqref{eq:dg-discrete-representation-solution}, the functions $\bm{\phi}_{j}(\mathbf{x}, p)$ are the basis vectors of $V_{h}$. The total number of DoFs is the same as the total number of basis functions, namely $N = \text{card}(\mathcal{T}_{h}) \dim([\mathbb{Q}^{k}(T)]^{n}) = \text{card}(\mathcal{T}_{h}) n (k + 1)^{d+1}$, where $\text{card}(\mathcal{T}_{h})$ is the number of cells in the triangulation $\mathcal{T}_{h}$.

As with all FEMs, we seek to construct a linear system to determine the DoF $\zeta_{j}(t)$. This is achieved by multiplying eq.~\eqref{eq:well-posed-advection-reaction-eq} with a basis function $\bm{\phi}_{i} \in V_{h}$ from the left, replacing $\mathbf{f}$ with its discrete counterpart $\mathbf{f}_{h} \in V_{h}$ and integrating the equation over the domain $D = \bigcup_{T \in \mathcal{T}_{h}} T$. This yields
\begin{equation}
  \label{eq:step-towards-linear-system}
  \sum_{T \in \mathcal{T}_{h}} \int_{T} \bm{\phi}_{i} \cdot \left(  \partial_{t}\mathbf{f}_{h}
  + (\mat{\beta} \cdot \tilde{\nabla}) \mathbf{f}_{h}
  + \mat{R} \mathbf{f}_{h} \right)
  = \sum_{T \in \mathcal{T}_{h}} \int_{T} \bm{\phi}_{i} \cdot \left(  \partial_{t}\bm{\phi}_{j}
  + (\mat{\beta} \cdot \tilde{\nabla}) \bm{\phi}_{j}
  + \mat{R} \bm{\phi}_{j} \right)\zeta_{j}
  = \sum_{T \in \mathcal{T}_{h}} \int_{T} \bm{\phi}_{i} \cdot \mathbf{s}
  \quad\forall i \in \{1, \dots, N\} \,.
\end{equation}
Note that the integration variables are implicit to improve the readability, i.e. we did not include $\mathrm{d}^{d+1}\xi$ in the integral expressions.

Keeping in mind that $\bm{\phi}_{i}$ and $\bm{\phi}_{j}$ are defined ``cell-wise'', eq.~\eqref{eq:step-towards-linear-system} yields a set of equations for each cell $T$, which is independent of the set of equations determining the coefficients $\zeta_{j}$ on the neighbouring cells. In the language of FEMs, the degrees of freedom on one cell are decoupled from the degrees of freedom on neighbouring cells. Physically, we expect a flux from one cell to the next, because we are solving an advection equation. Thus, we expect that the degrees of freedom of different cells \emph{do} couple and, hence, we adapt the linear system~\eqref{eq:step-towards-linear-system} such that it incorporates this expectation.

\subsection{Numerical flux}
\label{sec:numerical-flux}

The usual approach is to manipulate the linear system~\eqref{eq:step-towards-linear-system} in such a way that there is a flux from one cell to the next while taking care that the original equation is recovered if we use the exact solution $\mathbf{f}$ instead of its approximation $\mathbf{f}_{h}$, i.e. that the manipulated system is \textit{consistent} with the original problem.

To investigate the fluxes between the cells, we integrate the advection term $\mat{\beta} \cdot \tilde{\nabla} \mathbf{f}_{h}$ by parts and apply the divergence theorem, i.e.
\begin{equation}
  \label{eq:advection-term-by-parts}
  \begin{split}
    \sum_{T \in \mathcal{T}_{h}} \int_{T} (\bm{\phi}_{i})_{k} \mat{\beta}^{\alpha}_{kl}\tilde{\nabla}_{\alpha}(\,\mathbf{f}_{h})_{l}
     & = \sum_{T \in \mathcal{T}_{h}}  \int_{\partial T} n_{\alpha}\left((\bm{\phi}_{i})_{k} \mat{\beta}^{\alpha}_{kl}(\,\mathbf{f}_{h})_{l} \right)
    - \int_{T}(\bm{\phi}_{i})_{k} \tilde{\nabla}_{\alpha}\mat{\beta}^{\alpha}_{kl}(\,\mathbf{f}_{h})_{l}
    - \int_{T}\tilde{\nabla}_{\alpha}(\bm{\phi}_{i})_{k} \mat{\beta}^{\alpha}_{kl}(\,\mathbf{f}_{h})_{l}                                             \\
     & = \sum_{T \in \mathcal{T}_{h}}  \int_{\partial T} \bm{\phi}_{i} \cdot (\mathbf{n} \cdot \mat{\beta}) \mathbf{f}_{h}
    - \int_{T}\bm{\phi}_{i} \cdot (\tilde{\nabla} \cdot \mat{\beta})\mathbf{f}_{h}
    - \int_{T}  (\tilde{\nabla}\bm{\phi}_{i} \cdot \mat{\beta}) \mathbf{f}_{h}  \,,
  \end{split}
\end{equation}
where $\partial T$ denotes the surface of the cell $T$.

We define the vector $\mathbf{J} \coloneqq (\mathbf{n} \cdot \mat{\beta}) \mathbf{f}$ that has the physical interpretation of a flux in the direction of the normal $\mathbf{n}$. This becomes clear when we look at an arbitrary component, say $(\mathbf{J})_{i} = n_{\alpha} \mat{\beta}^{\alpha}_{ij}(\, \mathbf{f})_{j} \coloneqq n_{\alpha} \, \mathbf{j}^{\alpha}_{i}$. The matrices $\mat{\beta}^{\alpha}$ \lq\lq mix'' the components of $\mathbf{f}$, which results in a current density $\mathbf{j}_{i}$. This current density is projected onto the normal $\mathbf{n}$ and each component of $(\mathbf{J})_{i}$ is the projection of a different current density $\mathbf{j}_{i}$ onto $\mathbf{n}$.

\begin{figure}
  \centering
  \begin{tikzpicture}
    \draw (0,0) rectangle (2,2);
    \node[anchor=north west] at (0,2) {$T_{1}$};
    \node[anchor=north] at (2, 0) {$F$};
    \draw (2,0) rectangle (4,2) node[anchor=north east]{$T_{2}$};
    \draw[->, thick](2, 1) -- (2.75, 1) node[anchor=south] {$\mathbf{n}_{F}$};

    \node at (1.75, 1.35){$\mathbf{f}_{h,1}$};
    \node at (2.35, 0.6) {$\mathbf{f}_{h,2}$};
    \draw (0, 1) .. controls (1,1.75) .. (2,1.5);
    \draw (2, 1) .. controls (3, 0.5) .. (4,1.1);
  \end{tikzpicture}
  \caption{Two adjacent cells $T_{1}$ and $T_{2}$. The discrete representation of the solution
    $\mathbf{f}_{h}$ is not continuous on the cell interface.\cite[cf. Fig. 1.4]{Pietro_MathematicalAspectsDG}}
  \label{fig:interfaces}
\end{figure}

In a next step, we focus on the sum over the surface integrals in eq.~\eqref{eq:advection-term-by-parts} and, in particular, we will look at a single cell interface $F$ as the one depicted in Fig.~\ref{fig:interfaces}. An integral over such a cell interface $F$ consists in a contribution from cell $T_{1}$ and another contribution from cell $T_{2}$, namely
\begin{equation}
  \label{eq:flux-at-interface}
  \int_{F} \bm{\phi}_{i_{1}} \cdot (\mathbf{n}_{T_{1}} \cdot \mat{\beta}) \mathbf{f}_{h,1} + \bm{\phi}_{i_{2}} \cdot (\mathbf{n}_{T_{2}} \cdot \mat{\beta}) \mathbf{f}_{h,2}
  =   \int_{F} \bm{\phi}_{i_{1}} \cdot (\mathbf{n}_{F}\cdot \mat{\beta}) \mathbf{f}_{h,1} - \bm{\phi}_{i_{2}} \cdot (\mathbf{n}_{F} \cdot \mat{\beta}) \mathbf{f}_{h,2}
\end{equation}
The introduction of the subscript $1$ and $2$ reflects that the basis functions of $V_{h}$ are defined on each cell, i.e. there is a set of basis functions defined on $T_{1}$ and another one defined on $T_{2}$. Moreover, the outward normal $\mathbf{n}_{T_{1}} = \mathbf{n}_{F}$ and $\mathbf{n}_{T_{2}} = - \mathbf{n}_{F}$, see Fig.~\ref{fig:interfaces}. Note that the flux through $F$ is not unique, because $\mathbf{f}_{h}$ is discontinuous. Physically, we expect that $\mathbf{f}$ is continuous and, hence, that the flux is single-valued, which motivates the replacement of the two fluxes appearing in eq.~\eqref{eq:flux-at-interface} with a \textit{numerical flux} $\bm{\mathring{J}}_{F}(\mathbf{f}_{h,1}, \mathbf{f}_{h,2})$, which is a single-valued function of both values of $\mathbf{f}_{h}$.

To ensure consistency, we have to require that the numerical flux reduces to the physical flux, if we plug in the exact solution $\mathbf{f}$, i.e. $\bm{\mathring{J}}_{F}(\mathbf{f}, \mathbf{f}) = \mathbf{J} = (\mathbf{n} \cdot \mat{\beta}) \mathbf{f}$.

The introduction of the numerical flux changes the integral over the cell interface $F$ in eq.~\eqref{eq:flux-at-interface} to
\begin{equation}
  \label{eq:numerical-at-interface}
  \int_{F} \left(\bm{\phi}_{i_{1}}  - \bm{\phi}_{i_{2}}\right) \cdot \bm{\mathring{J}}_{F}(\mathbf{f}_{h,1}, \mathbf{f}_{h,2}) \coloneqq \int_{F} \llbracket \bm{\phi}_{i} \rrbracket \cdot \bm{\mathring{J}}_{F}(\mathbf{f}_{h,1}, \mathbf{f}_{h,2}) \,,
\end{equation}
where we defined the symbol $\llbracket \bm{\phi}_{i} \rrbracket$ to denote the jump of the basis functions at a cell interface.

We now replace the sum over the surface integrals in eq.~\eqref{eq:advection-term-by-parts} with a sum over the cell interface and boundary face integrals, namely
\begin{equation}
  \label{eq:sum-interface-boundaries}
  \sum_{T \in \mathcal{T}_{h}}  \int_{\partial T} \bm{\phi}_{i} \cdot (\mathbf{n} \cdot \mat{\beta}) \mathbf{f}_{h}
  \longrightarrow \sum_{F \in \mathcal{F}^{i}_{h}} \int_{F} \llbracket \bm{\phi}_{i} \rrbracket \cdot \bm{\mathring{J}}_{F}(\mathbf{f}_{h,1}, \mathbf{f}_{h,2})
  + \sum_{F \in \mathcal{F}^{b}_{h}} \int_{F} \bm{\phi}_{i} \cdot \bm{\mathring{J}}^{B}_{F}(\mathbf{f}_{h}) \,.
\end{equation}
We stress that this transition includes the introduction of the numerical flux, which is a deliberate manipulation of the linear system~\eqref{eq:step-towards-linear-system}. Furthermore, we introduced the sets $\mathcal{F}^{i}_{h}$ and $\mathcal{F}^{b}_{h}$ whose elements are the cell interfaces and the faces of the boundary cells respectively. $\bm{\mathring{J}}^{B}_{F}(\mathbf{f}_{h})$ denotes the numerical flux through the latter.

The manipulated linear system for the coefficients $\zeta_{j}$ is obtained through two replacements: Firstly, we replace the sum over the surface integrals in eq.~\eqref{eq:advection-term-by-parts} with the sum over the face integrals in eq.~\eqref{eq:sum-interface-boundaries}. Second, we replace the advection term in the original linear system~\eqref{eq:step-towards-linear-system} with the result of the previous replacement. This yields
\begin{equation}
  \label{eq:linear-system-degrees-of-freedom}
  \sum_{T \in \mathcal{T}_{h}} \int_{T}\bm{\phi}_{i} \cdot \partial_{t}\mathbf{f}_{h}
  + \int_{T} \bm{\phi}_{i} \cdot \left\{ \mat{R}   - (\tilde{\nabla} \cdot \mat{\beta})\right\} \mathbf{f}_{h}
  -  \int_{T} (\tilde{\nabla}\bm{\phi}_{i} \cdot \mat{\beta} ) \mathbf{f}_{h}
  + \sum_{F \in \mathcal{F}^{i}_{h}} \int_{F} \llbracket \bm{\phi}_{i} \rrbracket \cdot \bm{\mathring{J}}_{F}(\mathbf{f}_{h,1}, \mathbf{f}_{h,2})
  + \sum_{F \in \mathcal{F}^{b}_{h}} \int_{F} \bm{\phi}_{i} \cdot \bm{\mathring{J}}^{B}_{F}(\mathbf{f}_{h})
  = \sum_{T \in \mathcal{T}_{h}} \int_{T} \bm{\phi}_{i} \cdot \mathbf{s}
\end{equation}
for all $i \in \{1, \dots, N\}$. We note that this is a system of ordinary differential equations (ODEs) which determines the degrees of freedom $\zeta_{j}(t)$ of $\mathbf{f}_{h}$, see eq.~\eqref{eq:dg-discrete-representation-solution}. We call the system \textit{semi-discretised}, because it is discretised in space but not in time.

Whether the solution to the system of ODEs~\eqref{eq:linear-system-degrees-of-freedom} approximates the exact solution $\mathbf{f}$ depends on many things, inter alia, on the choice of the numerical flux $\bm{\mathring{J}}_{F}$ and on the time stepping method used. A possible choice for the numerical flux is an \textit{upwind flux}. If an \textit{explicit Runge--Kutta method} (ERK) of order two (or three) is used, it is necessary to make the assumption that the exact solution $\mathbf{f}$ and the source term $\mathbf{s}$ are smooth enough to show that the ERK method converges over time. We note that any explicit time stepping method only converges if the time step is chosen in agreement with a suitable \textit{CFL-condition}\footnote{Information on time stepping methods like ERK can, for example, be found in \cite[Chapter II]{Hairer_SolvingOrdinaryDifferentialEquations} and an explanation of the CFL-condition is given in \cite[Section 4.4]{LeVeque_FVM}}. We conclude that if we choose an upwind flux, together with an ERK method, and if the exact solution and the source term are smooth enough, the dG method converges in time and space to the exact solution, \cite[see][Lemma 7.27 and Lemma 7.28 and references therein]{Pietro_MathematicalAspectsDG}.

In general, there are many different choices for the numerical flux, a typical one for a system of equations is the \textit{(local) Lax--Friedrichs flux}, \cite[see for example][p. 204]{Cockburn_IntroductionToDGConvectionDominated}. However, if a problem is advection dominated, it makes sense to use this knowledge to determine a more precise numerical flux. An upwind flux does exactly this, cf.~\cite[Section 4.8]{LeVeque_FVM}.

The upwind flux is defined as
\begin{equation}
  \label{eq:definition-upwind-flux}
  \bm{\mathring{J}^{U}_{F}}(\mathbf{f}_{h,1}, \mathbf{f}_{h,2}) \coloneqq \mat{W}\left(\bm{\Lambda}_{+}\mat{W}^{\transpose}\mathbf{f}_{h,1} + \bm{\Lambda}_{-}\mat{W}^{\transpose}\mathbf{f}_{h,2}\right)
  \quad\text{with } (\mathbf{n}_{F} \cdot \mat{\beta}) \mat{W}= \mat{W} \bm{\Lambda} \text{ and } \bm{\Lambda} = \bm{\Lambda}_{+} + \bm{\Lambda}_{-} \,,
\end{equation}
where $\mat{W}$ and $\bm{\Lambda}$ are the eigenvectors and eigenvalues of the matrix $(\mathbf{n} \cdot \mat{\beta})$ respectively. This definition can, for example, be found in~\cite[Section 2.4]{Hesthaven_NodalDG}. The eigenvalue matrix $\bm{\Lambda}$ is split into two matrices, namely into $\bm{\Lambda}_{+}$ with positive eigenvalues and zeros on its diagonal and $\bm{\Lambda}_{-}$ with negative eigenvalues and zeros on its diagonal. $(\mathbf{n} \cdot \mat{\beta})$ is symmetric, $\mat{W}^{\transpose} = \mat{W}^{-1}$. The result of the product $\mat{W}^{\transpose}\mathbf{f}_{h,i}$ are \textit{the characteristic variables}. An insightful physical interpretation of the upwind flux is given in \cite[p. 47]{LeVeque_FVM}.

A local Lax--Friedrichs flux only needs the maximum eigenvalue of $\mathbf{n}_{F} \cdot \mat{\beta}$ whereas an upwind flux requires the diagonalisation of $n \times n$ matrices, where $n = (l_{\text{max}} + 1)^{2}$, at each interface in every time step. We note that the size of the matrices grows quadratically with the order of the spherical harmonic expansion and, if possible, it is best to avoid the computation of eigenvectors and eigenvalues of so large matrices.

However, if we restrict the triangulation $\mathcal{T}_{h}$ to (hyper-)rectangles, we are able to avoid the necessity to solve an eigenproblem at each interface. Considering that all normals of the faces of a rectangle can be parallel to the coordinate axes, the matrix $(\mathbf{n}_{F} \cdot \mat{\beta})$ simplifies. For example, if we are interested in the upwind flux through a face whose normal points in the $x$-direction, i.e. $\mathbf{n}_{F} = \mathbf{e}_{x}$, then
\begin{equation}
  \label{eq:normal-equal-e-x}
  (\mathbf{n}_{F} \cdot \mat{\beta}) = (\mathbf{e}_{x} \cdot \mat{\beta}) = \mat{\beta}^{x}
  = U^{x}\mat{1} + V \mat{A}^{x} \,,
\end{equation}
where we used the definition of $\mat{\beta}^{\alpha}$ given in eq.~\eqref{eq:definitions-advection-reaction-eq}.

Now, let $\mathbf{w}$ be an eigenvector of $\mat{A}^{x}$ with eigenvalue $\lambda$, then
\begin{equation}
  \label{eq:eigenvalues-eigenvectors-beta-x}
  \left(U^{x}\mat{1} + V \mat{A}^{x}\right) \mathbf{w}
  = U^{x}\mathbf{w} + \lambda V \mathbf{w} = (U^{x} + \lambda V) \mathbf{w} \,.
\end{equation}
Hence, $\mathbf{w}$ is \emph{also} an eigenvector of $\mat{\beta} \cdot \mathbf{e}_{x} = \mat{\beta}^{x} = U^{x}\mat{1} + V\mat{A}^{x}$, and the corresponding eigenvalue is $U^{x} + \lambda V$.

We conclude that we have to determine the eigenvectors and eigenvalues of $\mat{A}^{x}$ \emph{once} to get the upwind flux in the $x$-direction at all interfaces and at all times, because the eigenvectors of $\mat{\beta}^{x}$ do \emph{not} change and the eigenvalues can be updated by multiplying them with $V$ and adding $U^{x}$, see eq.~\eqref{eq:eigenvalues-eigenvectors-beta-x}.

The same is true for the upwind fluxes in $y$- and $z$-direction with the only difference that we do \emph{not} have to compute the eigenvalues and eigenvectors of $\mat{A}^{y}$ and $\mat{A}^{z}$. In \cite{Schween24a} we show that all three matrices have the same eigenvalues and that the eigenvectors of $\mat{A}^{y}$ and $\mat{A}^{z}$ can be computed by rotating the eigenvectors of $\mat{A}^{x}$.

The upwind fluxes in the momentum direction, i.e. in the $p$-direction, are more complicated, because
\begin{equation}
  \label{eq:beta-p}
  (\mathbf{n}_{p} \cdot \mat{\beta}) = \mat{\beta}^{p}
  = - \gamma m \frac{\mathrm{d} U_{a}}{\mathrm{d} t} \mat{A}^{a}
  - p \frac{\partial U_{b}}{\partial x^{a}} \mat{A}^{a}\mat{A}^{b}
\end{equation}
contains sums and products of the $\mat{A}^{a}$ matrices and the above arguments, which we used to avoid finding a solution to the eigenproblem, do not apply. Thus, we end up solving an eigenproblem at each interface whose normal points in the $p$-direction.

\subsection{Numerical flux at the boundaries of the domain}
\label{sec:numerical-flux-at-the-boundaries}

Having defined the numerical flux to be the upwind flux, we almost have an explicit form of the linear system~\eqref{eq:linear-system-degrees-of-freedom}, which determines the approximate solution $\mathbf{f}_{h}$. ``Almost'', because we have not yet defined the numerical flux $\bm{\mathring{J}}^{B}_{F}(\mathbf{f}_{h})$ through the boundary.

We note that the approximate solution $\mathbf{f}_{h}$ must fulfil the zero inflow boundary condition~\eqref{eq:zero-inflow-bc} and that the choice of $\bm{\mathring{J}}^{B}_{F}$ can enforce it. It is this idea that informs the definition of the boundary flux. Assume it was the upwind flux~\eqref{eq:definition-upwind-flux} as well and that $\mathbf{f}_{h}$ was single-valued on the boundary, i.e. $\mathbf{f}_{h} = \mathbf{f}_{h,1} = \mathbf{f}_{h,2}$, then the boundary flux would be
\begin{equation}
  \label{eq:perliminary-numerical-boundary-flux}
  \bm{\mathring{J}}^{B}_{F}(\mathbf{f}_{h})
  = \mat{W}\left(\bm{\Lambda}_{+}\mat{W}^{\transpose}\mathbf{f}_{h}
  + \bm{\Lambda}_{-}\mat{W}^{\transpose}\mathbf{f}_{h}\right) \,,
\end{equation}
and its second term would be the flux into the domain.

We now enforce zero inflow by setting
\begin{equation}
  \label{eq:zero-inflow-boundary}
  \mat{W} \bm{\Lambda}_{-}\mat{W}^{\transpose}\mathbf{f}_{h} = 0
  \iff \bm{\Lambda}_{-}\mat{W}^{\transpose}\mathbf{f}_{h} = 0  \,.
\end{equation}
Where $\bm{\Lambda}_{-} \mat{W}^{\transpose} \mathbf{f}_{h}$ ``picks out'' the inflow part of $\mathbf{f}_{h}$, because the multiplication with $\mat{W}^{\transpose}$ yields the characteristic variables and the multiplication with $\bm{\Lambda}_{-}$ eliminates all the characteristic variables which do \emph{not} contribute to the inflow. The reason being that the diagonal elements of $\bm{\Lambda}_{-}$ corresponding to outflow components are zero.

We formalise the ``picking-out'' of inflow components by introducing the matrix $\mat{1}_{-}$, which has ones where $\bm{\Lambda}_{-}$ has non-zero entries and zeros everywhere else. We use this matrix to define the inflow part of $\mathbf{f}$, i.e. $\mathbf{f}^{-} \coloneqq \mat{1}_{-}\mat{W}^{T}\mathbf{f}$.

An implication of setting the inflow components of $\mathbf{f}_{h}$ to zero is that the discrete solution $\mathbf{f}_{h}$ fulfils the boundary condition~\eqref{eq:zero-inflow-bc}. This motivates to define the boundary flux to be
\begin{equation}
  \label{eq:definition-boundary-flux}
  \bm{\mathring{J}}^{B}_{F}(\mathbf{f}_{h})
  \coloneqq \mat{W}\bm{\Lambda}_{+}\mat{W}^{\transpose}\mathbf{f}_{h} \,.
\end{equation}

We emphasise that we do \emph{not} prescribe values of $\mathbf{f}$ on the boundary to enforce the zero inflow boundary condition. It is implicit in the definition of $\bm{\mathring{J}^{B}_{F}}$ and every solution $\mathbf{f}_{h}$ to the linear system~\eqref{eq:linear-system-degrees-of-freedom} of ODEs is automatically in agreement with it. In the language of FEMs it is said that the boundary conditions are enforced \textit{weakly}\footnote{Weakly enforced boundary conditions do \emph{not} hold on every point on the boundary, they only hold \textit{almost everywhere}.}, which is typical for dG methods.

\subsection{Time stepping method}
\label{sec:time-stepping-method}

With the definition of the boundary flux, we have an explicit expression for all the terms in the system of ODEs~\eqref{eq:linear-system-degrees-of-freedom}, and we can now solve it using any of the standard time stepping methods.

In a first step, we bring the system of ODEs in a particularly simple form to ease the application of a time stepping method, i.e.
\begin{equation}
  \label{eq:matrix-formulation-system-odes}
  \mat{M} \frac{\mathrm{d}\bm{\zeta}}{\mathrm{d} t} + \mat{D}(t)\bm{\zeta} = \mathbf{h}(t) \,.
\end{equation}
Where the components of the vector $\bm{\zeta}$ are the degrees of freedom of the approximate solution $\mathbf{f}_{h}$, see eq.~\eqref{eq:dg-discrete-representation-solution}. Furthermore, we introduced the symbols
\begin{equation}
  \label{eq:definition-mass-and-dg-matrix}
  \begin{split}
    (\mat{M})_{ij}   & \coloneqq \sum_{T \in \mathcal{T}_{h}} \int_{T} \bm{\phi}_{i} \cdot \bm{\phi}_{j} \, ,                                                           \\
    (\mat{D})_{ij}   & \coloneqq \sum_{T \in \mathcal{T}_{h}} \int_{T} \bm{\phi}_{i} \cdot \left\{ \mat{R}   - (\tilde{\nabla} \cdot \mat{\beta})\right\} \bm{\phi}_{j}
    -  \int_{T}(\tilde{\nabla}\bm{\phi}_{i} \cdot \mat{\beta} ) \bm{\phi}_{j}
    + \sum_{F \in \mathcal{F}^{i}_{h}} \int_{F} \llbracket \bm{\phi}_{i} \rrbracket \cdot \bm{\mathring{J}}_{F}(\bm{\phi}_{j,1}, \bm{\phi}_{j,2})
    + \sum_{F \in \mathcal{F}^{b}_{h}} \int_{F} \bm{\phi}_{i} \cdot \bm{\mathring{J}}^{B}_{F}(\phi_{j}) \text{ and }                                                    \\
    (\mathbf{h})_{i} & \coloneqq \sum_{T \in \mathcal{T}_{h}} \int_{T} \bm{\phi}_{i} \cdot \mathbf{s} \,.
  \end{split}
\end{equation}

We apply the \textit{$\Theta$-method} to time step, namely
\begin{equation}
  \label{eq:theta-method}
  \mat{M}\frac{\bm{\zeta}^{n} - \bm{\zeta}^{n-1}}{\Delta t} =
  (1 - \Theta)\left(\mathbf{h}^{n-1} - \mat{D}^{n-1} \bm{\zeta}^{n-1}\right)
  + \Theta \left(\mathbf{h}^{n} - \mat{D}^{n}\bm{\zeta}^{n}\right) \, ,
\end{equation}
which allows one to switch between an implicit and an explicit time stepping. The superscript $n$ means that the respective quantity is evaluated at time step $n$, for example, $\bm{\zeta}^{n} \coloneqq \bm{\zeta}(n \Delta t)$. $\Theta$ takes values in the interval $[0, 1]$ and $\Theta = 0$ results in the \textit{forward (or explicit) Euler method}, $\Theta = 1$ gives the \textit{backward (or implicit)} Euler method and $\Theta = 1/2$ corresponds to the \textit{Crank--Nicholson method}.

The initial conditions $\bm{\zeta}^{0}$ for eq.~\eqref{eq:matrix-formulation-system-odes} are the coefficients of $\mathbf{f}_{h}(0) = \zeta_{j}(t = 0) \bm{\phi}_{j}$, i.e. the coefficients of the approximate solution at $t = 0$. We use the initial condition $\mathbf{f}(\mathbf{x}, p , 0) = \mathbf{f}^{0}$ for the exact solution to compute these coefficients, i.e. $\bm{\zeta}^{0}$. This is achieved by projecting the initial conditions onto the finite element space
\begin{equation}
  \label{eq:zeta-initial-condition}
  \sum_{T \in \mathcal{T}_{h}} \int_{T} \bm{\phi}_{i} \mathbf{f}_{h}(0)
  = \sum_{T \in \mathcal{T}_{h}} \int_{T} \bm{\phi}_{i} \cdot \bm{\phi}_{j} \zeta_{j}(t = 0)
  = (\mat{M})_{ij}(\bm{\zeta}^{0})_{j}
  = \sum_{T \in \mathcal{T}_{h}} \int_{T} \bm{\phi}_{i} \cdot \mathbf{f}^{0}  \,.
\end{equation}

We highlight that the $\Theta$-method is an implicit method for $\Theta > 0$, i.e. $\bm{\zeta}^{n}$ appears on both sides of eq.~\eqref{eq:theta-method}. We rearrange eq.~\eqref{eq:theta-method} for $\bm{\zeta}^{n}$ and arrive at the linear system
\begin{equation}
  \label{eq:system-matrix-rhs}
  \left(\mat{M} + \Delta t \Theta \mat{D}^{n}\right)\bm{\zeta}^{n}
  = \left(\mat{M} - \Delta t (1 - \Theta) \mat{D}^{n-1}\right)  \bm{\zeta}^{n-1}
  + \Delta t \left( (1 - \Theta) \mathbf{h}^{n-1} + \Theta\mathbf{h}^{n} \right) \,.
\end{equation}
This system is solved \emph{iteratively} in every time step.

We implemented the dG method and $\Theta$-method, as outlined in this section, in \sapphire using the finite element library \texttt{deal.ii} \cite{dealii2019, dealii95}. Moreover, \sapphire users can use an explicit fourth order Runge--Kutta method (ERK4).

\section{Tests and Simulations}
\label{sec:tests}

In this section we investigate the abilities of \sapphire in four examples, and eventually we apply it to a standard astrophysical scenario, namely we simulate the acceleration of particles at a parallel shock.

The four examples have been selected to showcase specific features of the code and to highlight its \emph{numerical} accuracy. In particular, the first test case shows that the dG space discretisation together with the various time-stepping methods converge as theoretically expected. The second test case investigates consequences of the truncation of the spherical harmonic expansion at finite order $l_{\text{max}}$. This is expanded on in the third example, which quantitatively investigates the effect of truncating the expansion at $l_{\text{max}}$. The last example, i.e. the simulation of diffusive shock acceleration at a parallel shock, shows that \sapphire is applicable to actual astrophysical scenarios.

Dimensionless units are used when solving the VFP equation~\eqref{eq:real-system-of-equations} in \sapphire. The definitions of the units and their reference values are given in Tab.~\ref{tab:sapphire-units}. Length and time are defined in terms of a reference gyroradius and a gyrofrequency, motivated by the fact that \sapphire is written with physical effects occurring on gyroscales in mind.

\begin{table}
  \centering
  \caption{Units and their reference values in \sapphire.}
  \label{tab:sapphire-units}
  \begin{tabular}{cclll}
    \hline
    Unit     & Definition                  & Reference                                                                     & Value                                          &                                 \\
    \hline
    $t^{*}$  & $t \underline{\omega}_{g} $ & $\underline{\omega}_{g} \coloneqq \underline{q} \underline{B}/ \underline{m}$ & 9.578833160\,$\cdot 10^{-3}$\, s$^{-1}$        &                                 \\
    $x^{*}$  & $x/\underline{r}_{g}$       & $\underline{r}_{g} \coloneqq \underline{m} c/ \underline{q}\underline{B}$     & 3.129738800\,$\cdot 10^{10}$ \,m               & 1.014279269$\cdot 10^{-6}$\, pc \\
    $p^{*}$  & $p/\underline{p}$           & $\underline{p} \coloneqq \underline{m} c $                                    & 5.014394376\,$\cdot 10^{-19}$\,kg\,m\,s$^{-1}$ & 938.272088\,MeV c$^{-1}$        \\
    $V^{*}$  & $ V / c$                    & $c$                                                                           & 2.997924580\,$\cdot 10^{8}$ \,m\,s$^{-1}$      &                                 \\
    $q^{*}$  & $q / \underline{q} $        & $\underline{q} \coloneqq e $                                                  & 1.602176634\,$\cdot 10^{-19}$\, C              &                                 \\
    $m^{*}$  & $m/\underline{m}$           & $\underline{m} \coloneqq m_{p} $                                              & 1.672621923\,$\cdot 10^{-27}$\,kg              & 938.272088\,MeV c$^{-2}$        \\
    $B^{*} $ & $B/\underline{B}$           & $\underline{B}$                                                               & 1.0\,$\cdot 10^{-10}$\,T                       & 1 $\mu$G                        \\
    \hline
  \end{tabular}
\end{table}

\sapphire is designed in a way that terms of the VFP equation can be included or excluded in the simulation as required. We apply the following naming scheme:
\begin{alignat}{2}
  & \partial_{t}\mathbf{f}                                                                                       & \quad & \text{(time-evolution term)} \notag  \\
  & {} + \left(U^{a}\mat{1} + V \mat{A}^{a}\right) \partial_{x^{a}}\mathbf{f}                                    &       & \text{(spatial advection term)}\notag \\
  \label{eq:vfp-terms-names}
  & {} - \left(\gamma m \frac{\mathrm{d} U_{a}}{\mathrm{d} t} \mat{A}^{a}
  + p \frac{\partial U_{b}}{\partial x^{a}} \mat{A}^{a}\mat{A}^{b}\right)\partial_{p} \mathbf{f}
  + \left(\frac{1}{V} \epsilon_{abc} \frac{\mathrm{d} U^{a}}{\mathrm{d} t} \mat{A}^{b}\mat{\Omega}^{c}
  + \epsilon_{bcd} \frac{\partial U_{b}}{\partial x^{a}}\mat{A}^{a}\mat{A}^{c}\mat{\Omega}^{d}\right) \mathbf{f} &       & \text{(momentum term)}               \\
  & {} - \omega_{a}\mat{\Omega}^{a}\mathbf{f}                                                                    &       & \text{(rotation term)}\notag        \\
  & {} + \nu \mat{C}\mat{f}                                                                                      &       & \text{(collision term)} \notag        \\
  & {} = \mathbf{s} \,.                                                                                          &       & \text{(source term)} \notag
\end{alignat}
We emphasise that it is possible to solve the above equation for different configuration space dimensions, namely for $\mathbf{x} \in \mathbb{R}^{d}$ with $d = 1, 2$. If the momentum terms are deactivated, i.e. if monoenergetic particles\footnote{Up to now, \sapphire can only simulate up to three dimensions of the \emph{reduced phase-space}, $\bm{\xi}$. If the momentum term is deactivated, the reduced phase space is equivalent to configuration space $\bm{\xi} = \bm{x}^T$, i.e. simulations in up to three physical space dimensions are possible.} are simulated $d$ can equal $3$. Additionally, we allow to choose between a linear momentum variable $p$ and a logarithmic momentum variable $\ln p$. The momentum terms are adapted accordingly.

The first two examples retain the time-evolution, the spatial advection term and the rotational term. The third example includes the time-evolution, the spatial advection and the collision term. In the last example all terms are included in the simulation.

\subsection{Convergence study}
\label{sec:convergence-study}

In this example we demonstrate the numerical accuracy of the dG space discretisation and four different time-stepping methods by simulating a simple test-case whose exact solution we present in the next subsection. Moreover, we study how this accuracy changes with different time steps, cell sizes and the polynomial degree of the dG basis functions described in Section~\eqref{sec:discrete-representation-solution}.

\paragraph{Description}
\label{sec:description-convergence-study}
We consider a mono-energetic distribution of particles in a static background plasma ($\mathbf{U} =0$) that is permeated with a magnetic field ($\mathbf{B} = B_{0} \mathbf{e}_{z}$) with no scattering between the particles and the plasma ($\nu = 0$). This amounts to neglecting the collision, momentum and source terms in eq.~\eqref{eq:vfp-terms-names}. We allow only spatial derivatives in the $x$-direction reducing the spatial advection term. In this case the system of equations~\eqref{eq:real-system-of-equations} reduces to

\begin{equation}
  \label{eq:system-of-equations-convergence-study}
  \partial_{t}\mathbf{f} + V \mat{A}^{x} \partial_{x}\mathbf{f}
  - \omega_{z} \mat{\Omega}^{z}\mathbf{f}
  = 0 \,.
\end{equation}

Physically, the system models a distribution function $f$, homogeneous in $y$ and $z$, that describes charged particles gyrating about $\mathbf{B}$ in the $x-y$ plane.

To arrive at an analytic solution, we consider a toy model and truncate at $l_{\text{max}} = 1$. Equation~\eqref{eq:system-of-equations-convergence-study} then becomes

\begin{equation}
  \begin{split}
     & \partial_{t} f_{000} +
    \frac{V}{\sqrt{3}} \partial_{x} f_{100}
    = 0                              \\
     & \partial_{t} f_{110} -
    \omega_{z} f_{100}
    = 0                              \\
     & \partial_{t} f_{100} +
    \frac{V}{\sqrt{3}} \partial_{x} f_{000} +
    \omega_{z} f_{110}
    = 0                              \\
     & \partial_{t} f_{111} = 0 \, .
  \end{split}
  \label{eq:convergence-study-explicit-equations}
\end{equation}

This set of equations can be combined to an equation for $f_{100}$, i.e.
\begin{equation}
  \partial_{t}^2 f_{100} -
  \frac{V^2}{3} \partial_{x}^2 f_{100} +
  \omega^2_{z} f_{100}
  = 0 \, ,
\end{equation}
which for initial conditions $f_{100}(t=0) = f_{110}(t=0) = 0$ and periodic boundary condition in a box with length $L$, has a separable solution, namely
\begin{equation}
  f_{100}(t, x) = \sum^{\infty}_{n = 0}  c_n \sin\left(\sqrt{\omega^2_{z} + c_n^2}\, t\right)
  \left[ A_{n} \sin(k_n x) + B_{n} cos(k_n x) \right] \, .
\end{equation}
Here, we introduced the wave number $k_n = 2 \pi n/L $ and $c_n=Vk_n/\sqrt{3}$ with $n \in \mathbb{N}$. Combining these conditions with the $\partial_t f_{100}$ equation in \eqref{eq:convergence-study-explicit-equations}, a consistent solution can be found for the initial condition:
\begin{equation}
  \label{eq:convergence-study-initial-condition-f000}
  f_{000}(t=0, x) = \sum^{\infty}_{n = 0} \sqrt{\omega^2_{z} + c_n^2}
  \left[ A_n \cos(k_n x) - B_n \sin(k_n x) \right] \,.
\end{equation}

It follows that
\begin{align}
  \label{eq:convergence-study-analytical-solution-isotropic-part}
  f_{000}(t,x)  & = \sum^{\infty}_{n=0} \frac{c_n^2}{\sqrt{\omega^2_{z} + c_n^2}}
  \left[
  \cos\left(\sqrt{\omega^2_{z} + c_n^2} t\right) - 1 +
  \frac{\omega^2_{z} + c_n^2}{c_n^2}
  \right]
  \left[ A_n \cos(k_n x) - B_n \sin(k_n x) \right]                                         \\
  f_{110}(t,x)  & = \sum^{\infty}_{n=0} \frac{\omega_{z} c_n}{\sqrt{\omega^2_{z} + c_n^2}}
  \left[ 1 - \cos\left(\sqrt{\omega^2_{z} + c_n^2} t\right) \right]
  \left[ A_n \sin(k_n x) + B_n \cos(k_n x) \right]                                         \\
  \label{eq:convergence-study-analytical-solution-100-part}
  f_{100}(t, x) & = \sum^{\infty}_{n=0} c_n \sin\left(\sqrt{\omega^2_{z} + c_n^2} t\right)
  \left[ A_n \sin(k_n x) + B_n \cos(k_n x) \right] \, .
\end{align}

\paragraph{\sapphire setup}
We emphasise that the solution presented in eq.~(\ref{eq:convergence-study-analytical-solution-isotropic-part}-\ref{eq:convergence-study-analytical-solution-100-part}) is the mathematical solution to the reduced system of equations~\eqref{eq:convergence-study-explicit-equations}, and not the physical solution that one could in principle determine, for example via Liouville's theorem. This allows for a direct comparison with numerical solutions determined with \sapphiren, including the time-evolution, spatial advection and rotation terms. To match the analytic solution, the dimension of the configuration space is $\text{dim}(\xi) = 1$, and the expansion order is set to $l_{\text{max}} = 1$. The numerical value of the $\mathbf{B}$-field is chosen to be $\mathbf{B}^{*}= 2\pi \mathbf{e}_{z}$. Here, the asterisk means that the quantities are given in the units described in Tab.~\ref{tab:sapphire-units}. We fix the energy of the particles to $\gamma = 2$, implying that $\omega^{*}_{z} = \pi$. To ensure positivity of $f_{000}$, we consider for the initial condition $A_0=2$ and $B_1 = 1$, all other $A_n, B_n$ being set to zero. The size of the box is $L^{*} = 20$.

Since this is a one dimensional example, the computational grid (or mesh) is a line and in all the computed cases. The cells have the size $h = \Delta x^{*} = L^{*}/N_{\text{cells}}$, where $N_{\text{cells}} \in \mathbb{N}$ is the number of cells. Simulations are run until $t^{*}_{F} = 10/\omega^{*}_{z}$.

\paragraph{Results}

\begin{figure}
  \centering
  \resizebox{0.48\textwidth}{!}{
    \centering
    \begin{tikzpicture}[baseline]
      \begin{loglogaxis}[
          axis lines=left,
          title={Convergence in $\Delta t$},
          xlabel={$\Delta t^{*}$},
          ylabel={max rel. $L^{2}$-error},
          xmin=1e-3, xmax=0.2,
          ymin=1e-13, ymax=2e-2,
          grid = none,
          legend entries={ERK4, FE, BE,CN},
          cycle list name=exotic,
          legend pos=south east,
        ]
        \addplot+ [solid, mark=*, thick] table[x index=0, y index=1, col sep=comma] {errors_FE.csv};
        \addplot+ [solid, mark=square*, thick] table[x index=0, y index=1, col sep=comma] {errors_BE.csv};
        \addplot+ [solid, mark=triangle*, thick] table[x index=0, y index=1, col sep=comma] {errors_CN.csv};
        \addplot+ [solid, mark=diamond*, thick] table[x index=0, y index=1, col sep=comma] {errors_ERK4.csv};

        \addplot [domain=1e-3:0.2, dashed, semithick] {3e-1 * x};
        \node [gray, anchor=south west] at (1e-3,5e-4) {$\Delta t$};
        \addplot [domain=1e-3:0.2, dashed, semithick] {3e-1 * x^2};
        \node [gray, anchor=south west] at (1e-3,5e-7) {$\Delta t^2$};
        \addplot [domain=1e-3:0.2, dashed, semithick] {3e-1 * x^4};
        \node [gray, anchor=south west] at (1e-3,1e-12) {$\Delta t^4$};
      \end{loglogaxis}
    \end{tikzpicture}
  }
  \hskip 15pt
  \resizebox{0.48\textwidth}{!}{
    \centering
    \begin{tikzpicture}[baseline]
      \begin{loglogaxis}[
          axis lines=left,
          title={Convergence in $\Delta x$},
          ylabel={max rel. $L^{2}$-error},
          xmin=0.15625, xmax=2.5,
          ymin=1e-10, ymax=2,
          grid = none,
          legend entries={$k=1$, $k=2$, $k=3$, $k=4$, $k=5$},
          cycle list name=exotic,
          legend pos=north west,
          legend columns=2,
          ylabel={max rel. $L^{2}$-error},
          xlabel={$\frac{\Delta x^{*}}{L^{*}}$},
          xtick={2.5, 1.25, 0.625, 0.3125, 0.15625},
          xticklabels={$\frac{1}{8}$, $\frac{1}{16}$, $\frac{1}{32}$, $\frac{1}{64}$, $\frac{1}{128}$},
        ]
        \addplot+ [solid, mark=*, thick] table[x index=0, y index=1, col sep=comma] {errors_k1.csv};
        \addplot+ [solid, mark=square*, thick] table[x index=0, y index=1, col sep=comma] {errors_k2.csv};
        \addplot+ [solid, mark=triangle*, thick] table[x index=0, y index=1, col sep=comma] {errors_k3.csv};
        \addplot+ [solid, mark=diamond*, thick] table[x index=0, y index=1, col sep=comma] {errors_k4.csv};
        \addplot+ [solid, mark=pentagon*, thick] table[x index=0, y index=1, col sep=comma] {errors_k5.csv};

        \addplot [domain=0.1:3, dashed, semithick] {4e-3 * x^2};
        \node [gray, anchor=south west] at (1,6e-3) {$\Delta x^2$};
        \addplot [domain=0.1:3, dashed, semithick] {1e-4 * x^3};
        \node [gray, anchor=south west] at (1,2e-4) {$\Delta x^3$};
        \addplot [domain=0.1:3, dashed, semithick] {2e-6 * x^4};
        \node [gray, anchor=south west] at (1,4e-6) {$\Delta x^4$};
        \addplot [domain=0.1:3, dashed, semithick] {3e-8 * x^5};
        \node [gray, anchor=south west] at (1,7e-8) {$\Delta x^5$};
        \addplot [domain=0.1:3, dashed, semithick] {5e-10 * x^6};
        \node [gray, anchor=south west] at (1,2e-9) {$\Delta x^6$};
      \end{loglogaxis}
    \end{tikzpicture}
  }
  \caption{Left: Convergence in $\Delta t^{*}$ for different time-stepping methods with $\Delta x^{*} = L^{*}/64$ and $k=5$. The CFL condition is violated for $\Delta t^{*} > t_{\text{CFL}} \approx 10^{-2}$. Right: Convergence in $\Delta x^{*}$ for different polynomial degrees $k$ with ERK4 and $\Delta t^{*} = 10^{-2}$.}
  \label{fig:convergence-study}
\end{figure}
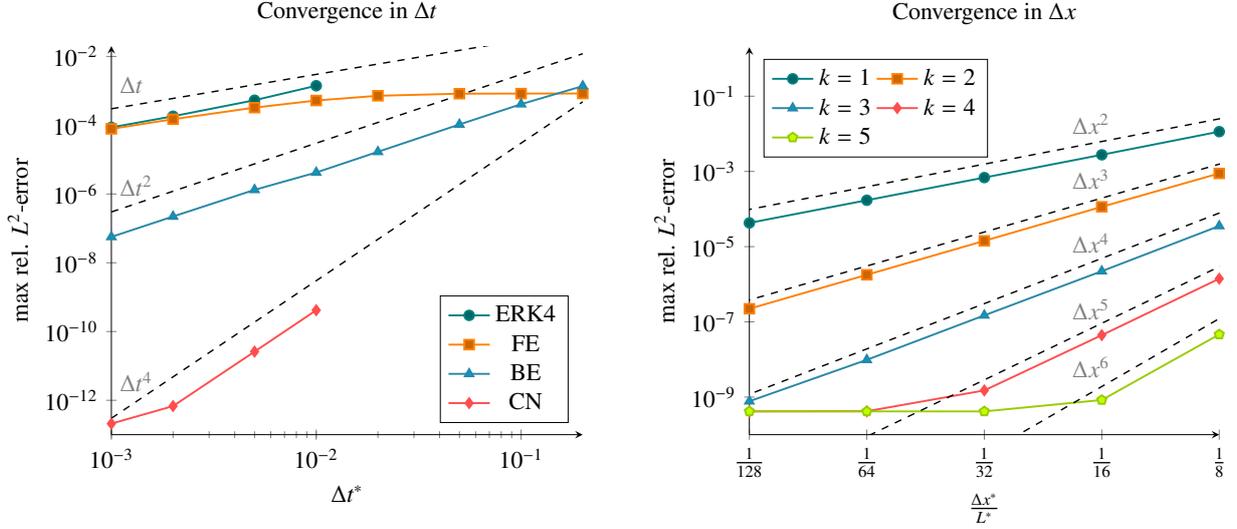

In this example we are interested in quantitatively comparing the numerical solution to the analytical solution. We restrict ourselves to the isotropic part $f_{000}$ of the distribution function $f$ and, thus, introduce the $L_{2}$ norm of the error of $f_{000}$, i.e.
\begin{equation}
  \label{eq:definition-l2-norm-error}
  \left\| f^{n}_{000,h} - f_{000}(t^n) \right\|_{L^{2}}
  \coloneqq \left(\int_{L^{*}} \left| f^{n}_{000,h}(x^{*}) - f_{000}(x^{*},t^{n}) \right|^{2} \mathrm{d}x^{*}\right)^{1/2} \,,
\end{equation}
where $f^{n}_{000,h}(x^{*})$ is the numerical approximation of the solution at time step $t^{n}$. This means that we do \emph{not} integrate the error over time. Instead, we introduce the maximum relative error,
\begin{equation}
  \text{max rel. $L^2$ error} \coloneqq \max_{t^{n}} \frac{\left\| f_{000, h}^{n} - f_{000}(t^n) \right\|_{L^2}}{\left\| f_{000, h}^{n} \right\|_{L^2}} \,,
\end{equation}
where the maximum is that of all time steps.

The left plot in Fig.~\ref{fig:convergence-study} demonstrates how the error changes when we reduce the time step $\Delta t^{*}$ for a fixed spatial resolution $\Delta x^{*}$, comparing the different time-stepping methods implemented in \sapphire. Each data point corresponds to one simulation run. In all these simulations, we use a high spatial resolution of $\Delta x^{*} = L^{*}/64 = 0.3125$ with polynomial degree $k=5$. This ensures that the numerical error of the time-stepping methods is larger than the spatial discretisation error.

The explicit fourth order Runge Kutta (ERK4) and forward Euler (FE) methods are only shown for time steps respecting the following CFL condition (see for example \cite[Sec. 2.3.3]{Cockburn_IntroductionToDGConvectionDominated} and \cite[Sec. 3.1.4]{Pietro_MathematicalAspectsDG}):
\begin{equation}
  \Delta t^{*} \leq t_{\text{CFL}} \approx \frac{1}{2k+1} \frac{\Delta x^{*}}{\beta^{*}_{\text{max}}} \,.
\end{equation}
$\beta^{*}_{\text{max}} = U^{*} + \lambda_{\text{max}} V^{*}$ is the maximum velocity of the spatial advection term, with $\lambda_{\text{max}}$ the maximum eigenvalue of the $\mat{A}^{x}$ matrix. In this example $\beta^{*}_{\text{max}} = V^{*}/\sqrt{3}$.

The error associated to the ERK4 and FE methods scale as $\Delta t^4$ and $\Delta t$ respectively, as expected. For ERK4 the spatial error dominates when $\Delta t^{*} \approx 10^{-3}$, and the error plateaus. The error of the Crank--Nicolson (CN) method scales as $\Delta t^2$, while the implicit backward Euler (BE) method is only first order accurate $\Delta t$ for small time steps, though the error saturates for large $\Delta t$ due to the boundedness of the analytical solution: Large errors of the time stepping method cause the numerical solution and the analytical solution to be represented by cosines of different frequencies. Thus, the error is bounded by the amplitude of the cosines.

The right-hand plot in Fig. \ref{fig:convergence-study} shows the convergence with respect to the spatial resolution $\Delta x^{*}$ for different polynomial degrees $k$. For these simulations we used the ERK4 method with a fixed time step $\Delta t^{*} = 10^{-2}$, respecting the CFL condition and ensuring that the time stepping error is subdominant ($\text{max rel. $L^2$ error} \sim 10^{-9}$). The error is therefore dominated by the spatial discretisation error up to very high spatial resolution. As expected this error scales as $\Delta x^{k+1}$ \cite[Sec. 2.2.4]{Cockburn_IntroductionToDGConvectionDominated}.\footnote{If the initial condition would not belong to the Sobolev space $H^{k+2}$ but only to $H^{k+1}$, the error would scale as $\Delta x^{k+1/2}$.}

\subsection{Advection in a constant magnetic field}
\label{sec:gyro-advection}

As we apply a spectral method, a truncation of the spherical harmonic series expansion can result in a discrepancy between the physical solution and the numerical solution found. In this example we once more consider a test case for which the physical solution is known precisely and demonstrate how truncating the expansion at a finite $l_{\text{max}}$ affects the solution.

\paragraph{Description}

The test scenario studied in this paragraph is very similar to the one of the previous example, i.e. we compute the distribution of charged and monoenergetic particles moving in a background plasma with a constant magnetic field $\mathbf{B} = B_0 \mathbf{e}_{z}$ in which the particles are not scattered $(\nu = 0)$. We consider two cases, one where the background plasma is static ($\mathbf{U} = 0$) and another where it is moving at a constant velocity $\mathbf{U} = U_0 \left(\mathbf{e}_{x} + \mathbf{e}_{y}\right)$. This example is 2D; $\bm{\xi}=(x,y)^T$. The corresponding system of equations is
\begin{equation}
  \label{eq:gyro-advection-example-system-of-equations}
  \partial_{t}\mathbf{f} + \left(U_{0}\mat{1} +  V\mat{A}^{x}\right)\partial_{x}\mathbf{f} + \left(U_{0}\mat{1} +  V\mat{A}^{y}\right)\partial_{y}\mathbf{f}
  - \omega_{z} \mat{\Omega}^{z}\mathbf{f}
  = 0 \,.
\end{equation}

Neglecting scattering, the particle trajectories are known, and the solution can be readily found from Liouville's theorem, i.e. $\mathrm{d} f/\mathrm{d} t = 0$. Determination of the expansion coefficients however requires a numerical procedure which we prefer to avoid. We instead exploit the fact that the particles are constrained to gyrate about the magnetic field, such that the distribution function must return to its initial condition after one gyroperiod $T_{g}=2 \pi/\omega_g = 2 \pi \gamma m/q $. If the background plasma moves with constant, uniform velocity, the distribution function is translated in this time by a distance $|\mathbf{U}| T_{g} = \sqrt{2} U_{0} T_{g}$ in the direction of $\mathbf{U}$. We exploit this when comparing the numerical solution with the physical expectation at multiples of $T_{g}$.

\begin{figure}
  \centering
  \includegraphics[width=0.3\textwidth]{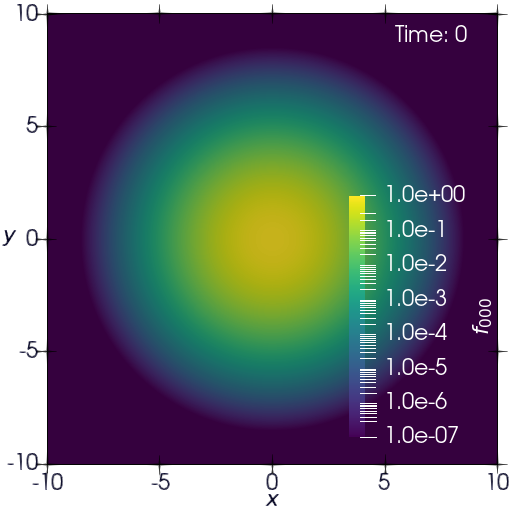}
  \includegraphics[width=0.3\textwidth]{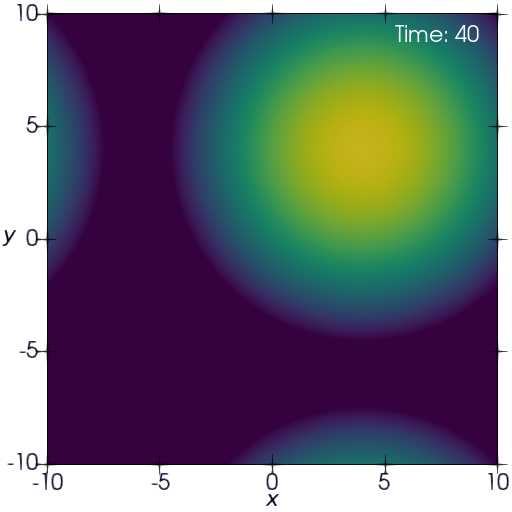}
  \includegraphics[width=0.3\textwidth]{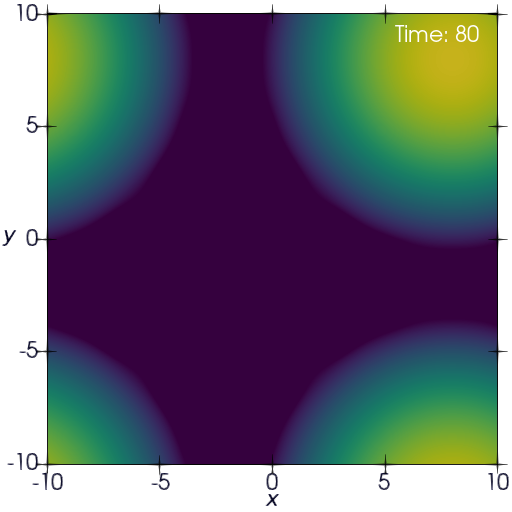}
  \includegraphics[width=0.3\textwidth]{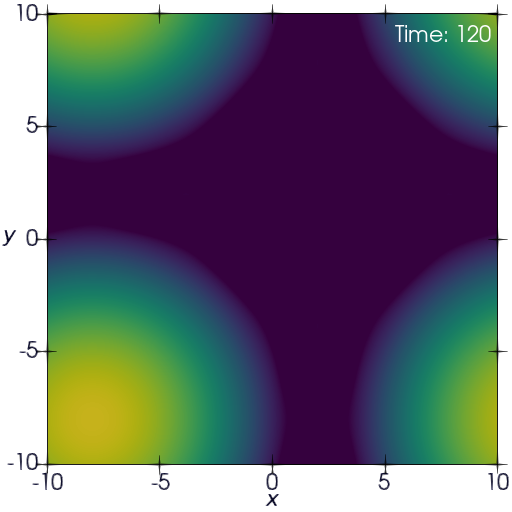}
  \includegraphics[width=0.3\textwidth]{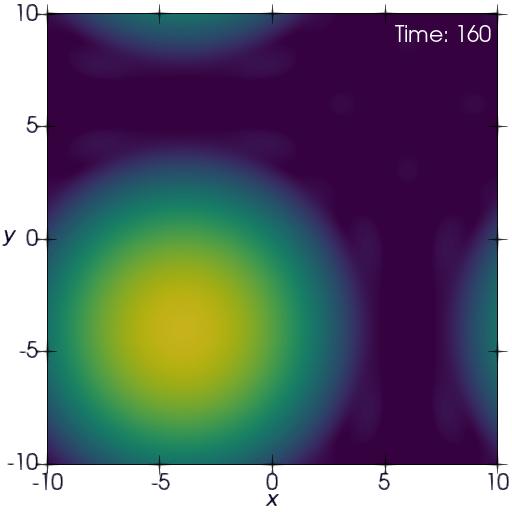}
  \includegraphics[width=0.3\textwidth]{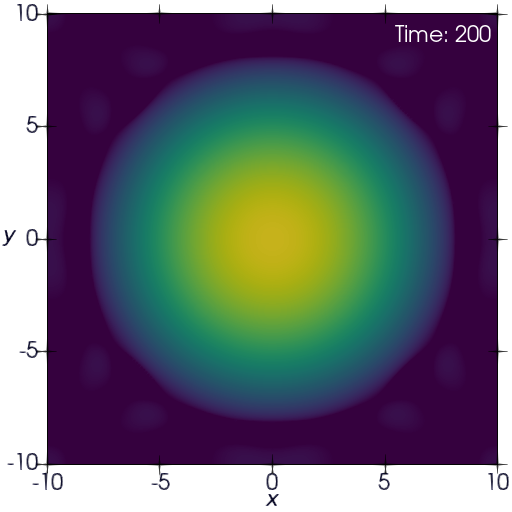}
  \caption{Time series showing the advection of the isotropic part of the distribution function $f_{000}$ gyrating in a constant magnetic field. We use $l_{\text{max}} = 3$, $\Delta x^{*} = 20/64$,  $k=3$ and a ERK4 time stepping method with $\Delta t^{*}=0.02$.}
  \label{fig:gyro-advection_timeseries}
\end{figure}

\paragraph{\sapphire setup}
In the computation of the numerical solution, we include the time-evolution term, the spatial advection term and the rotation term. As the gyration and advection in this example are restricted to the $x$--$y$-plane, we set the dimension of the configuration space to $d = 2$, i.e. $\bm{\xi} \in \mathbb{R}^{2}$. We truncate the expansion either at $l_{\text{max}} = 3$ or $l_{\text{max}} = 5$ to show how numerical solution converges with increasing spectral resolution.

We choose, in numerical units, the following parameters $U^{*}_0 = 0.1$, $B^{*}_0 = 2\pi$ and the energy of the particle is set by $\gamma = 2$. As an initial condition we choose a Gaussian for the isotropic part of the distribution function:
\begin{equation}
  f_{000}(t^{*}=0, \mathbf{x}^{*}) = \exp\left(- \frac{x^{*2} + y^{*2}}{2 \sigma^{*2}}\right) \,.
\end{equation}
All other expansion coefficients are set to zero. The standard deviation is $\sigma^{*} = 1.5$, equivalent to approximately five gyroradii.

The computational domain is a periodic box of size $L^{*} = 20$ that is uniformly refined such that $\Delta x^{*} = L^{*}/64$. We use polynomials of degree $k=3$ in conjunction with a ERK4 time stepping method. The time step size is $\Delta t^{*}=0.02$. We picked the spatial and temporal resolution such that the dominating error is produced by the cut-off in $l_{\text{max}}$.

\paragraph{Results}

In Fig.~\ref{fig:gyro-advection_timeseries} we show the advecting isotropic part of the distribution function at different time steps for a moving background plasma. We note that if we use a periodic box of length $L^{*} = N U^{*}_0 T^{*}_g$ with $N \in \mathbb{N}$ in conjunction with the prescribed constant velocity, the distribution function will return to its initial position after $N$ gyroperiods. In agreement with this consideration, we choose the parameters such that the particles described by the distribution function will gyrate $N = 100$ times before returning to their initial position at $t^{*} = 200$, see the lower right plot in Fig.~\ref{fig:gyro-advection_timeseries}.

\begin{figure}
  \centering
  \includegraphics[width=0.3\textwidth]{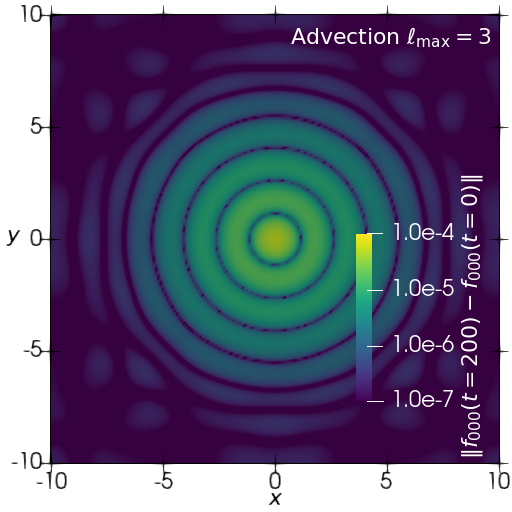}
  \includegraphics[width=0.3\textwidth]{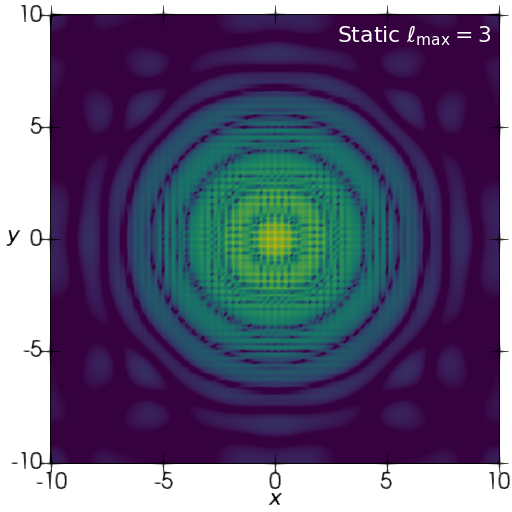}
  \includegraphics[width=0.3\textwidth]{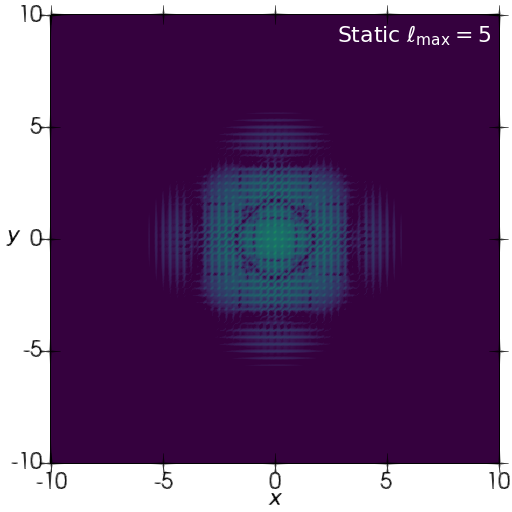}
  \caption{Residual for the different cases. Left: advection ($U_0=0.1$) with $l_{\text{max}} = 3$, Middle: static ($U^{*}_0=0$) with $l_{\text{max}} = 3$, Right: static ($U^{*}_0=0$) with $l_{\text{max}} = 5$. Spatial resolution and time step are the same as in Fig. \ref{fig:gyro-advection_timeseries}.}
  \label{fig:gyro-advection_residual}
\end{figure}

In Fig.~\ref{fig:gyro-advection_residual} the initial condition is compared with the result at $t^{*}=200$, by plotting the residual
\begin{equation}
  \label{eq:advection-example-error}
  \left|f_{h}(\mathbf{x}^{*}, t^{*}=200) - f_{h}(\mathbf{x}^{*}, t^{*}=0)\right| \,.
\end{equation}
Note that the residual is computed with the discrete representation of the initial condition, i.e. $f^{0}_{h}(\mathbf{x}^{*})$, and not with $f(t^{*}=0, \mathbf{x}^{*})$. For the case of the advecting isotropic distribution, we only plot the results for $l_{\text{max}} = 3$, whereas in the static case we compare $l_{\text{max}} = 3$ with $l_{\text{max}} = 5$.

The latter shows that truncating the expansion of the distribution function at $l_{\text{max}} = 3$ results in a solution which deviates slightly from the expected result, which should match the initial condition. The deviation is greatly reduced for larger $l_{\text{max}}$. The characteristic ring patterns are also a consequence of the truncation. Physically, the only frequency in the example is the gyrofrequency. However, a truncation at a finite $l_{\text{max}}$ introduces more frequencies, cf. the factor $c_{n}$ of the analytical solutions of the previous example given in eq.~\eqref{eq:convergence-study-analytical-solution-isotropic-part}. The difference in frequency leads to the interference pattern that is shown. The fact that we see rings is due to the axial symmetry of the (numerical and analytical) solution.

\subsection{Closure}
\label{sec:closure}
In the last example we investigated the effects of truncating the spherical harmonic expansion on the error, i.e. on the difference between an actual solution and its representation in terms of a finite series of spherical harmonics. We note that truncating the spherical harmonic expansion at $l_{\text{max}}$ closes the system of PDEs~\eqref{eq:real-system-of-equations}. The purpose of this example is to qualitatively discuss what physically justifies such a closure and to provide a heuristic for when a high-order expansion is needed.

\paragraph{Description}
\label{sec:description-closure}

As we will show in the following, the more spherical harmonics are included in the expansion of $f$, the better it can ``resolve'' anisotropic distributions of particles. What it means to resolve an anisotropic distribution can be understood by means of looking at the extremes: The extreme case of an anisotropic distribution is a beam, i.e. all particles move into the same direction. A beam could be thought of as the opposite of an isotropic particle distribution, i.e. in every direction moves the same amount of particles. Since $\theta$ and $\varphi$ encode the particles' direction of motion, a way to graphically illustrate anisotropies is to plot the average phase space density as a heatmap on the sphere. In this representation a beam is a bright point on the sphere and an isotropic distribution is a sphere of a single hueless colour. The number of spherical harmonics included in the expansion determines the angular resolution available to capture features in these heatmaps. A single point on the sphere requires an infinite resolution and, hence, beam-like distributions need a high-order expansion, i.e. a large $l_{\text{max}}$, whereas almost isotropic distributions can be resolved with small $l_{\text{max}}$. This way of thinking about anisotropic particle distributions also sheds light on our choice of the collision operator $C = \nu \Delta_{\theta, \varphi}/2$, see eq.~\eqref{eq:vfp-mixed-coordinates}: The directions of motion of the particles, given through $\theta$ and $\varphi$, are ``diffusing''. This, for example, means that the point representing a beam will smear out; the frequent collisions change the particles' directions of motion. Hence, scattering limits the scale on which anisotropic features appear on the sphere, reducing the number of spherical harmonics that need to be included in the expansion.

To demonstrate this, we set up the following one-dimensional example: We start off with an isotropic distribution of monoenergetic particles, homogeneous in $y$--$z$--plane with a Gaussian profile in $x$ direction. This distribution will evolve according to the following differential equation
\begin{equation}
  \label{eq:closure-example-f-complete}
  \frac{\mathrm{d} f}{\mathrm{d} t} = \frac{\partial f}{\partial t} + V_{x}\frac{\partial f}{\partial x} = \frac{\nu}{2} \Delta_{\theta,\varphi} f \,
\end{equation}
with the initial condition $$ f(t=0,x) = \frac{1}{\sqrt{2 \pi} \sigma} \exp\left(-\frac{x^{2}}{2\sigma^{2}}\right)\, , $$ where the standard deviation is set to $\sigma = 1$. Since we explore the effects of scattering only at the end of this example, we start with $\nu = 0$. Because all particles have the same energy, we have to fix their Lorentz factor. We choose $\gamma = 2$, which is equivalent to setting the magnitude of their velocities to $V = \sqrt{3}/2$. We note that the projected velocity along the $x$--axis depends on $\theta$, $V_{x} = V \cos\theta$.

\begin{figure}
  \centering
  \includegraphics[width=\linewidth]{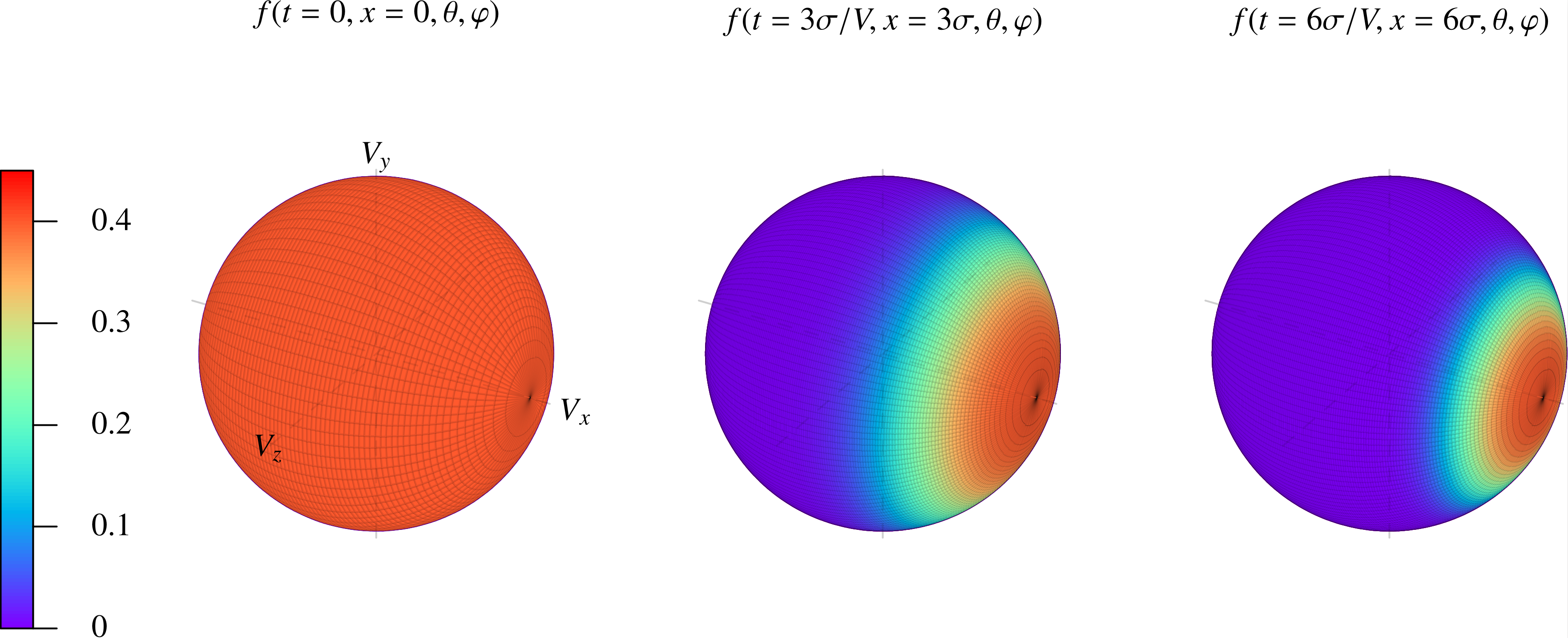}
  \caption{Analytical velocity distribution of the particles at three different space and time points.}
  \label{fig:closure-example-analytical-heatmaps}
\end{figure}

The solution to eq.~\eqref{eq:closure-example-f-complete}, keeping in mind that $\nu = 0$, can be computed with the method of characteristics and is
\begin{equation}
  \label{eq:closure-example-f-complete-solution}
  f(t,x) = \frac{1}{\sqrt{2 \pi} \sigma} \exp\left(-\frac{(x - V \cos\theta t)^{2}}{2\sigma^{2}} \right) \,.
\end{equation}
Particles with values of $\theta$ close to zero or $\pi$ move faster along the $x$-axis than particles with $\theta$ values around $\pi/2$. This leads to a separation of particles with different directions of motion, encoded by $\theta$. This is demonstrated in Fig.~\ref{fig:closure-example-analytical-heatmaps}.
We plot a heatmap of the average particle number density at $(x=0,t=0)$, at $(x=3 \sigma, t = 3 \sigma/ V = 6/\sqrt{3})$ and $(x=6 \sigma, t= 6 \sigma /V = 12/\sqrt{3})$. We choose the points such that a particle with velocity $V_{x} = \sqrt{3}/2$ starting at $x = 0$ has reached three or six standard deviations $\sigma$ respectively. In the second and third plot the separation of particles with different $\theta$ shows up as a hot spot at the pole of the sphere. The spot gets narrower the further out, and later we look at the distribution of particles. We now compare the solution in Fig. \ref{fig:closure-example-analytical-heatmaps} at $(x=6 \sigma, t= 6 \sigma /V = 12/\sqrt{3})$ with Sapphire++ using different $l_{\text{max}}$ and, moreover, we explore how scattering changes the particle distribution.

\paragraph{\sapphire setup}

The system of partial differential equations corresponding to eq.~\eqref{eq:closure-example-f-complete} is
\begin{equation}
  \label{eq:closure-example-f-expansion}
  \partial_{t}\mathbf{f} + V\mat{A}^{x}\partial_{x}\mathbf{f}   = -\nu \mat{C}\mathbf{f} \,,
\end{equation}
i.e. we include the time-evolution term, the spatial advection term and the collision term. The dimension of the configuration space is $d = 1$.

\begin{figure}
  \centering
  \includegraphics{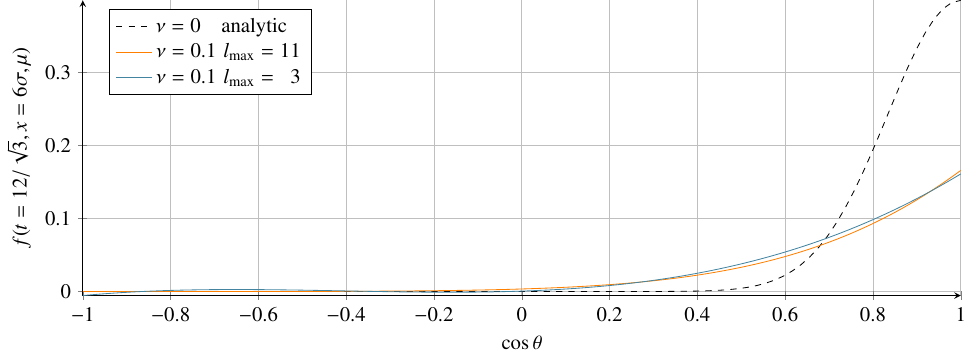}%
  \caption{ A comparison of the velocity distribution at $t = 12/\sqrt{3}$ and $x = 6 \sigma$ with and without scattering. If particles are scattered, a lower angular resolution (lower $l_{\text{max}}$) is sufficient to reproduce it.}
  \label{fig:closure-example-scattering-effects}
\end{figure}

The Lorentz factor of the particles is set to $\gamma = 2$ and, for our comparison with Fig.~\ref{fig:closure-example-analytical-heatmaps}, the scattering frequency is set to $\nu^{*} = 0$. In our exploration of the consequences of collisions for the anisotropies, we choose $\nu^{*} = 0.1$, see Fig.~\ref{fig:closure-example-scattering-effects}. The initial conditions for the system of PDEs~\eqref{eq:closure-example-f-expansion} is computed by projecting $f(t^{*} = 0, x^{*})$ onto the spherical harmonic space, namely
\[
  f_{000}(t^{*} = 0, x^{*}) = \int Y_{000} f(t^{*} = 0, x^{*}) \, \mathrm{d} \Omega
  = \frac{\sqrt{2}}{\sigma^{*}} \exp\left(-\frac{x^{*2}}{2\sigma^{*2}}\right) \,.
\]

The domain is $D = [-15., 15.]$, the cell size $\Delta x^{*} = 30/256$ and the polynomial degree $k =2$. The time step is $\Delta t^{*} = 1/(100 \sqrt{3})$ and the final time is $t^{*}_{F} = 12.$

We evaluate the expansion coefficients $f_{lms}$ and reconstruct the distribution function $f$ using eq.~\eqref{eq:spherical-harmonic-exp}. We compute the values of $f$ using 150 equally spaced points in the intervals $\cos\theta \in [-1, 1]$ and $\phi \in [0, 2\pi)$.

\begin{figure}
  \centering
  \includegraphics[width=\linewidth]{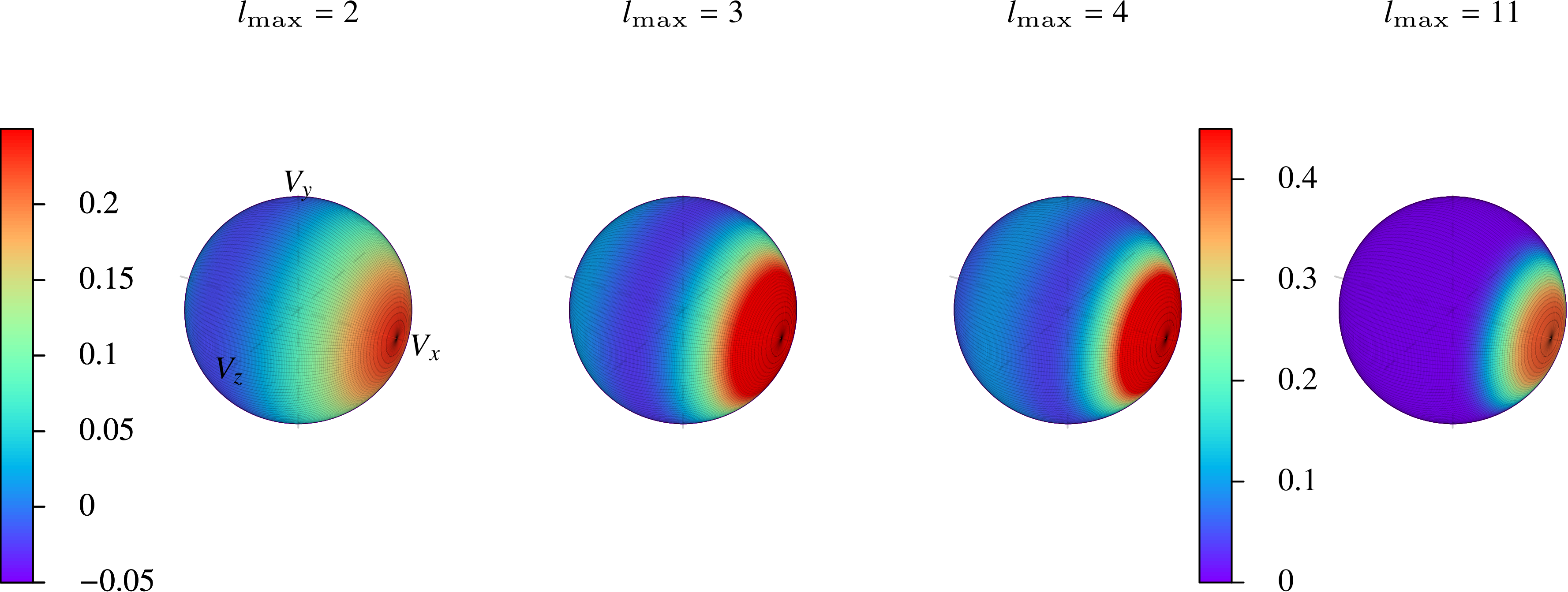}%
  \vspace{0.5cm} %
  \includegraphics{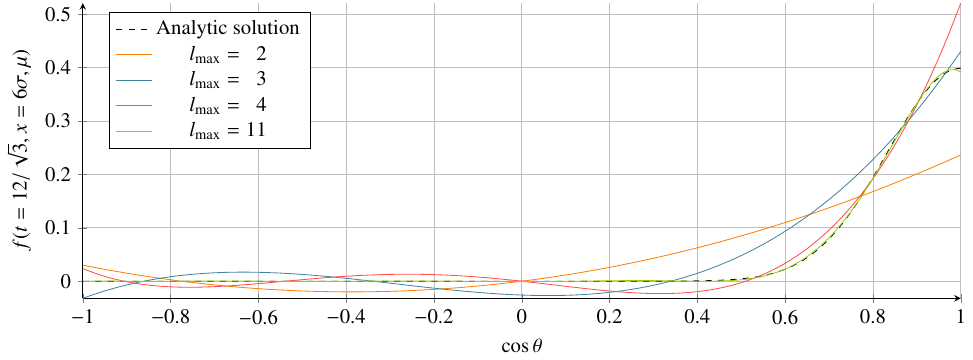}%
  \caption{Numerical velocity distribution of the particles at $t = 12/\sqrt{3}$ and $x = 6 \sigma$ for different $l_{\text{max}}$. $l_{\text{max}} = 11$ reproduces the analytical solution. }
  \label{fig:closure-example-numerical-heatmaps}
\end{figure}

\paragraph{Results}
Fig.~\ref{fig:closure-example-numerical-heatmaps} shows the angular distribution of the numerically computed distribution function $f$ for different $l_{\text{max}}$, at $t = 12/\sqrt{3}$ and $x = 6 \sigma$. In the upper panel we see qualitatively how increasing the expansion order $l_{\text{max}}$ captures the hot spot at the pole of the analytical solution shown on the right of Fig.~\ref{fig:closure-example-analytical-heatmaps}. In the lower panel, we show the same trend more quantitatively. We plot $f$ as a function of $\cos\theta$ for different $l_{\text{max}}$. We highlight that a low expansion order leads to a negative phase-space distribution of the particles and that this is a clear indicator for an insufficient truncation order. At $l_{\text{max}} = 11$, the difference between the numerical solution and analytical solution is visually almost indistinguishable.

In Fig.~\ref{fig:closure-example-scattering-effects} we illustrate how scattering changes the bright spot at the pole. The dashed line shows again the analytical solution $f$ for the case of zero scattering. Not having an analytical solution for $\nu > 0$, we ran a simulation with $\nu = 0.1$ using $l_{\text{max}} = 11$, which as demonstrated in Fig.~\ref{fig:closure-example-numerical-heatmaps} ensures an adequate angular resolution. Looking at the orange line, we see that the bright spot at the pole becomes broader. A second simulation run with $l_{\text{max}} = 3$, exemplifies that if scattering is present a lower expansion order is able to reproduce the salient feature.

\subsection{Diffusive Shock Acceleration at a Parallel shock}
\label{sec:parallel-shock}

\begin{figure}
  \centering
  \begin{tikzpicture}
    \draw[->] (0,0) -- (8,0) node[anchor=south] {\(x\)};
    \draw[thick, dashed] (4,-0.1) -- (4,2) node[anchor=south] {Shock};
    \node at (4,-0.3) {\(0\)};

    \draw[->, thick] (1,1.5) -- (3,1.5);
    \node[anchor=south] at (2.,1.5) {\(\mathbf{U}_{\text{shock}}\)};
    \draw[->, thick] (5.5,1.5) -- (6.,1.5);
    \node[anchor=south] at (5.5,1.5) {\(\mathbf{U}_{\text{shock}}/r\)};

    \draw[->, thick, gray, dashed] (2,0.5) -- (3,0.5);
    \node[anchor=south, gray] at (2,0.5) {\(\mathbf{B}\)};
    \draw[->, thick, gray, dashed] (5,0.5) -- (6,0.5);
    \node[anchor=south, gray] at (5,0.5) {\(\mathbf{B}\)};

    \draw[<-, thick] (3.5,0.5) -- (4.,0.5) node[anchor=south east] {\(\mathbf{n}\)};

    \node[anchor=south] at (0,2) {Upstream};
    \node[anchor=south] at (8,2) {Downstream};
  \end{tikzpicture}
  \caption{A parallel shock in the shock rest frame. $\mathbf{U}$ is the velocity of the background plasma. $\mathbf{B}$ is the mean magnetic field and $\mathbf{n}$ is the normal vector of the shock. We use $r$ to represent the compression ratio of the shock.}\label{fig:parallel-shock}
\end{figure}
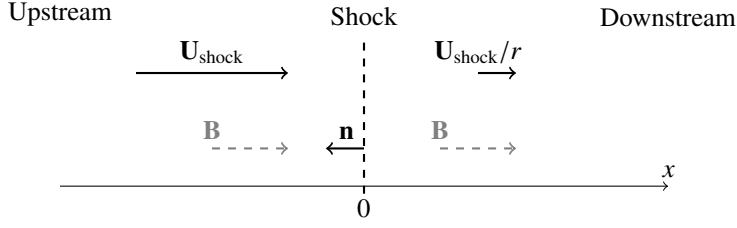

In this example we use \sapphire to simulate the time dependent acceleration of charged particles at a \textit{parallel shock} front, i.e. one in which the ambient magnetic field $\mathbf{B}$ is aligned with the direction of the shock's propagation, see Fig.~\ref{fig:parallel-shock}. The results of the simulation are compared to the \emph{steady-state} solution and an approximate time-dependent solution. The description given and the results presented in this example are taken from \cite{Drury1983} and \cite{Drury1991}, and we refer the reader to these works for further details.

\paragraph{Description}
\label{sec:derivation-analytic-solution}

The model we adopt makes the simplifying assumptions of an infinitely planar shock with both the magnetic field $\mathbf{B}$ and the velocity field of the background plasma $\mathbf{U}$ aligned with the shock normal $\mathbf{n}$. The coordinate system is chosen such that the $x$-axis is parallel to the shock normal. These assumptions imply that the $\mathbf{B}$-field is not modified by the shock. The distribution function $f$ should, thus, be independent of both the $y$ and $z$ coordinates, as well as the particles' gyrophase $\varphi$, because only a change in the $\mathbf{B}$-field could account for such a dependence. The parallel shock scenario is thus modelled in a reduced three-dimensional phase space, i.e. $\mathbb{R}^{3} = (x, p, \theta)$.

The velocity field $\mathbf{U}$ is assumed to undergo an infinitesimally narrow jump, i.e.
\begin{equation}
  \label{eq:discontinuous-velocity-profile}
  U(x) = \begin{cases}
    U_1 & \text{for } x <  0   \\
    U_2 & \text{for } x \geq 0
  \end{cases}\,,
\end{equation}
where $U_1=U_{\text{shock}}$ and $U_2=U_{\text{shock}}/r$, with $r$ denoting the \textit{compression ratio} of the shock, see Fig.~\ref{fig:parallel-shock}.

Furthermore, if all spatial variations of the distribution function are on large scales compared to the scattering mean free path, i.e. if the scattering frequency is high, it is sufficient to take only the first two terms in the expansion of the distribution function i.e. $l_{\text{max}} = 1$:
\begin{equation}
  \label{eq:f-l-max-1}
  F(x, p, \theta, t) = f_{000}(x,p,t)Y_{000} + f_{100}(x,p,t)Y_{100} = f(x,p,t) + a(x,p,t) \cos\theta
\end{equation}
where
\begin{equation}
  f \coloneqq \frac{1}{\sqrt{4\pi}} f_{000}
  \quad \text{and}\quad a\coloneqq \sqrt{\frac{3}{4\pi}} f_{100}\,
\end{equation}
are respectively the \emph{isotropic} and \emph{anisotropic} parts of the distribution. Note we use $F$ instead of $f$ to denote the ``full'' distribution function. We change our notation in this example to allow for a direct comparison with \cite[e.g. eq. (2.35)]{Drury1983}.

The steady-state solution for the isotropic part of the phase-space density is
\begin{alignat}{2}
  \label{eq:solution-1D-steady-transport-upstream}
  f(x,p_{1}) & =  f_{1}(p_{1})
  \exp\left(\int^{x}_{0} \frac{U(x')}{\kappa(x', p_{1})} \mathrm{d}x'\right)
             &                 & \quad  \text{ for } x < 0                               \\
  \label{eq:solution-1D-steady-transport-downstream}
  f(x,p_{2}) & = f_{2}(p_{2})  &                           & \quad \text{ for } x \geq 0
\end{alignat}
where $\kappa(x,p_{i}) = \lambda(x,p_{i}) V_{i}/3$ is \textit{the spatial diffusion coefficient} and $\lambda(x, p_{i}) = V_{i}/\nu$ is the particle \textit{scattering mean free path}. The subscripts $1$ and $2$ of the momentum variables are a consequence of the mixed-coordinate system, as we also use in \sapphiren, i.e. $p_{1}$ is measured in the upstream \emph{rest frame} and $p_{2}$ in the downstream~\cite[eq. (2.34)]{Drury1983}.

The high scattering frequency entails that the anisotropic part of the distribution function is much smaller than the isotropic part, $|a|\ll f$. In steady state it can be shown that (see~\cite[eq. (2.35)]{Drury1983})
\begin{equation}
  \label{eq:dipole-anisotropy}
  a(x,p_{i}) \approx - \lambda \frac{\partial f(x,p_{i})}{\partial x} \,,
\end{equation}
It follows from eq.~\eqref{eq:dipole-anisotropy}), that $a(x>0,p_{2}) = 0$, and $|a(x, p_{1})| = 3 (U_{1}/V_{1}) f(x, p_{1})$ for $ x < 0$. Hence, the dipole anisotropy vanishes in the downstream, while in the upstream it is of order $\mathcal{O}(U_{1}/V_{1})$ consistent with the initial assumption that $|a| \ll f$.

The steady-state spectrum of the isotropic component measured at the shock, i.e. $f_{i}(p_{i})$, can be computed by transforming the momenta to the shock rest frame, using $p'=p_i(1- U_i/V_i \cos\theta)$, expanding to first order in $U/V$, and matching the isotropic and anisotropic parts. Particles are assumed to be injected at $x=0$ and at a constant rate $Q$ with injection momentum $p_{0}$. The result of the computation is
\begin{equation}
  \label{eq:particle-spectrum-shock-rest-frame}
  f_{0}(p) = \frac{3Q}{p_{0}(U_{1}- U_{2})} \left(\frac{p}{p_{0}}\right)^{-3U_{1}/(U_{1}- U_{2})}
  = \frac{Q}{p_{0} U_{1}} \frac{3r}{r - 1} \left(\frac{p}{p_{0}}\right)^{-3r/(r - 1)}\,,
\end{equation}
where we used the subscript $0$ to highlight its dependence on the injection momentum.~\cite[eq. (3.24)]{Drury1983}.\footnote{We note that eq.~(3.24) in \cite{Drury1983} neglects a factor $1/p_{0}$. The solution to eq.~(3.21) ibid. should include it.} We emphasise here that despite the matching conditions being derived with momentum $p$ defined in the shock rest frame, to order $U/V$ the isotropic components must satisfy $f_0(p) = f_1(p_1) = f_2(p_2)$.

For later comparison with \sapphire, we can thus summarise that the steady-state distribution function in the vicinity of a parallel shock at which monoenergetic and isotropic particles are injected at a constant rate is
\begin{alignat}{2}
  \label{eq:full-distribution-function-upstream}
  F(x, p_{1}, \theta_{1}) & = \frac{Q}{p_{0}U_{1}} \frac{3r}{r - 1} \left(\frac{p_{1}}{p_{0}}\right)^{-3r/(r - 1)}
  \exp\left(\frac{3U_{1} \nu}{V_{1}^{2}} x\right)\left[1 - \frac{3 U_{1}}{V_{1}} \cos\theta_{1}\right]
                          &                                                                                        & \quad  \text{ for } x < 0                                    \\
  \label{eq:full-distribution-function-downstream}
  F(x, p_{2})             & = \frac{Q}{p_{0}U_{1}} \frac{3r}{r - 1} \left(\frac{p_{2}}{p_{0}}\right)^{-3r/(r - 1)} &                           & \quad \text{ for } x \geq 0  \,.
\end{alignat}
Note that in going from eq.~\eqref{eq:solution-1D-steady-transport-upstream} to eq.~\eqref{eq:full-distribution-function-upstream} we made the assumption that the scattering frequency $\nu$ does not depend on $x$.

Up to now, we concentrated on the steady-state solution. We are also interested in investigating the temporal evolution of the particle spectrum and how it compares to the numerically computed one. An \emph{approximate} analytic expression for the time-dependent spectrum at the shock has been given previously as: \cite[Sec. 3]{Drury1991}:
\begin{equation}
  \label{eq:approximate-temporal-evolution-spectrum}
  f(t^{*}, x^{*} = 0, p^{*}_{1}) = f_{0}(p^{*}_{1}) \phi(t^{*}) = f_{0}(p^{*}_{1}) \int^{t^{*}}_{0} \zeta(t') \mathrm{d}t' \,,
\end{equation}
whereas above, $f$ is the isotropic part of the distribution, and $\zeta(t^{*})$ can be understood as the acceleration time distribution at $x =0$ for acceleration from initial momentum $p^{*}_{0}$ to momentum $p^{*}_{1}$. For the case we consider it is
\begin{align}
  \label{eq:temporal-probability-distribution}
  \zeta(t^{*}) & = \frac{1}{\sqrt{2 \pi c_{2}}} \left(\frac{t^{*}}{c_{1}}\right)^{-3/2} \exp\left(\frac{-c_{1}(t^{*}- c_{1})^{2}}{2 t^{*} c_{2}}\right) \quad \text{ and accordingly}        \\
  \label{eq:cumulative-temporal-distribution}
  \phi(t^{*})  & = \frac{1}{2}\left[\exp\left(\frac{2 c^{2}_{1}}{c_{2}}\right) \erfc\left(\sqrt{\frac{c^{3}_{1}}{2 t^{*} c_{2}}} + \sqrt{\frac{c^{\phantom{3}}_{1} t^{*}}{2 c_{2}}}\right) +
    \erfc\left(\sqrt{\frac{c^{3}_{1}}{2 t^{*} c_{2}}} - \sqrt{\frac{c^{\phantom{3}}_{1}t^{*}}{2 c_{2}}}\right)\right] \,,
\end{align}
where $c_1$ and $c_2$ are the first two cumulants of the acceleration time distribution. The first cumulant $c_{1}$ corresponds to the mean acceleration time and is \cite[see][eq. (3.31)]{Drury1983}:
\begin{equation}
  \label{eq:mean-acceleration-time-shock-rest-frame}
  \begin{split}
    c_{1} \coloneqq t^{*}_{\text{acc}} \coloneqq \underline{\omega}_{g} \langle t \rangle
     & = \frac{3 \underline{\omega}_{g}}{U_{1} - U_{2}} \int^{p_{1}}_{p_{0}}\left(\frac{\kappa_{1}}{U_{1}} + \frac{\kappa_{2}}{U_{2}}\right) \frac{\mathrm{d} p}{p}
    =  \frac{r}{2 U^{*2}_{1} \nu^{*}} \frac{r + 1}{r - 1} \ln\left(\frac{1 + p^{*2}_{1}}{1 + p^{*2}_{0}}\right)\,,
  \end{split}
\end{equation}
where we used $\kappa_{1} = \kappa_{2} = p^{2}/(3 m^{2} \gamma^{2} \nu)$ and took the energy independent scattering frequency out of the integral.

Using the same diffusion coefficient, the second cumulant $c_{2}$, which is the variance of the acceleration time, is given by \cite[see][eq. (3.32)]{Drury1983}:
\begin{equation}
  \label{eq:variance-acceleration-time}
  \begin{split}
    c_{2} \coloneqq \sigma^{*2}_{\text{acc}}
    = \underline{\omega}^{2}_{g}\left(\langle t^{2} \rangle - \langle t \rangle^{2}\right)
     & = \frac{6\underline{\omega}^{2}_{g}}{U_{1} - U_{2}} \int^{p_{1}}_{p_{0}} \left(\frac{\kappa^{2}_{1}}{U^{3}_{1}} + \frac{\kappa^{2}_{2}}{U^{3}_{2}}\right) \frac{\mathrm{d} p}{p} \\
     & = \frac{1}{3\nu^{*2}} \frac{r}{U^{*4}_{1}} \frac{r^{3} + 1}{r - 1}
    \left[ \frac{1}{1 + p^{*2}_{1}} - \frac{1}{1 + p^{*2}_{0}} + \ln\left(\frac{1 + p^{*2}_{1} }{1 + p^{*2}_{0}}\right)\right] \,.
  \end{split}
\end{equation}

We note that the analytic expression for the temporal evolution of the spectrum given in eq.~\eqref{eq:approximate-temporal-evolution-spectrum}--\eqref{eq:cumulative-temporal-distribution} is exact if the diffusion coefficients are momentum independent, and satisfy $ \kappa_{1}/U^{2}_{1} = \kappa_{2}/U^{2}_{2}$ \cite{Toptyghin1980}. However, in \cite[Sec. 4]{Forman1983} it was pointed out that $\zeta(t^{*})$ could be used as an approximation to a general acceleration time distribution, i.e. for arbitrary diffusion coefficients, if the mean acceleration time and its variance are computed using the formulas eq.~\eqref{eq:mean-acceleration-time-shock-rest-frame} and \eqref{eq:variance-acceleration-time}. Moreover, it is required that $\zeta(t^{*})$ is normalised to unity\footnote{We numerically integrated $\zeta(t^{*})$ using the cumulants $c_{1}$ and $c_{2}$ as given in the text and found that its normalisation is correct within the errors of the integration method used.}, see also \cite[Sec. 3]{Drury1991}. Since our diffusion coefficient $\kappa$ does depend weakly on $p$ at low momenta, we expect the time-dependent spectrum in eq.~\eqref{eq:approximate-temporal-evolution-spectrum} to merely approximate the true time dependence.

\paragraph{\sapphire setup}
\label{sec:sapphire-setup}

For pragmatic reasons, the shock is modelled as a narrow transition of finite thickness, represented by a tanh profile for the velocity $\mathbf{U}$, and the point injection of the particles at the shock is approximated with a Gaussian. In this sense, the setup in this example does not match exactly the equations used to derive the analytical solution given in eq.~\eqref{eq:full-distribution-function-upstream} and \eqref{eq:full-distribution-function-downstream}.

In the simulation all terms are included, i.e. the time-evolution, the spatial advection, the momentum, rotation and collision terms, though the rotation term is not expected to contribute to the solution. The dimension of the configuration space is set to $d = 1$. Since the momentum terms are included, the reduced phase space is $\bm{\xi}=(x,p)^T$. As explained in the description of the example, it is sufficient to truncate the expansion at $l_{\text{max}} = 1$. The resulting PDE system~\eqref{eq:real-system-of-equations} consists of four equations for the expansion coefficients $f_{000}, f_{100}, f_{110}$ and $f_{111}$. Since we restrict the simulation to one spatial dimension and choose $\mathbf{B}$ and $\mathbf{U}$ to be aligned with the $x$-axis, the equations for $f_{110}$ and $f_{111}$ decouple and if the coefficients are initially zero, they remain so. Thus, the only equations containing non-zero terms are\footnote{The $\mathbf{B}$-field does not appear, because the non-zero elements of $\mat{\Omega}_{x}$ correspond to $f_{110}$ and $f_{111}$.}
\begin{equation}
  \label{eq:actual-soe-for-sapphire}
  \partial_{t}
  \begin{pmatrix}
    f_{000} \\
    f_{100}
  \end{pmatrix}
  +
  \begin{pmatrix}
    U                  & \frac{V}{\sqrt{3}} \\
    \frac{V}{\sqrt{3}} & U
  \end{pmatrix}
  \partial_{x}
  \begin{pmatrix}
    f_{000} \\
    f_{100}
  \end{pmatrix}
  - \frac{\partial U}{\partial x}
  \begin{pmatrix}
    \frac{p}{3}                & \frac{\gamma m}{ \sqrt{3}} \\
    \frac{\gamma m}{ \sqrt{3}} & \frac{3 p}{5}
  \end{pmatrix}
  \partial_{p}
  \begin{pmatrix}
    f_{000} \\
    f_{100}
  \end{pmatrix}
  +
  \begin{pmatrix}
    0 & -\frac{2}{\sqrt{3} V} \frac{\partial U}{\partial x} \\
    0 & \nu - \frac{2}{5} \frac{\partial U}{\partial x}
  \end{pmatrix}
  \begin{pmatrix}
    f_{000} \\
    f_{100}
  \end{pmatrix}
  =
  \begin{pmatrix}
    s_{000} \\
    0
  \end{pmatrix} \,,
\end{equation}
where a source term has been included on the right-hand side to represent the injection of the particles (assumed to be isotropic). We note all quantities in the above equation are dimensionless, see definitions in Tab.~\ref{tab:sapphire-units}.

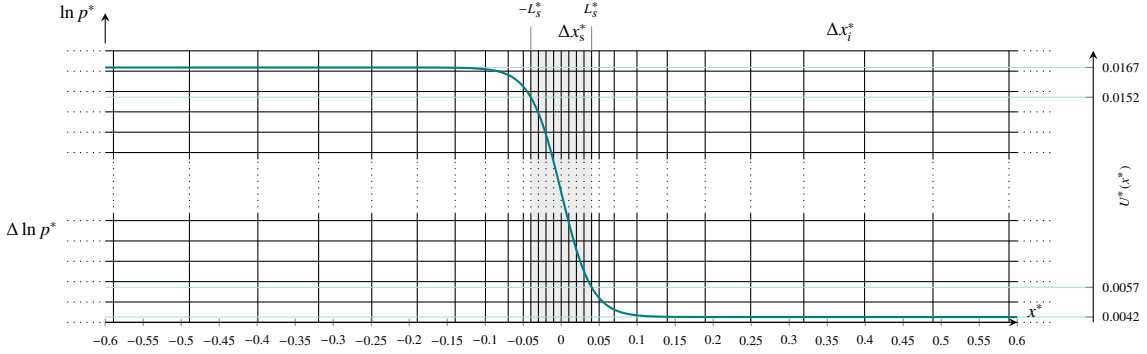
\begin{figure}
  \centering
  \begin{tikzpicture}
    \newcommand{\gridHeight}{3 * 1.2}
    \newcommand{\gridWidth}{5 * 1.2}
    \newcommand{\factor}{10}

    \newcommand{\shockWidth}{0.04}
    \newcommand{\stepSizeShock}{0.01}
    \pgfmathsetmacro\secondShockStep{-\shockWidth + \stepSizeShock}

    \newcommand{\numberpLines}{5}
    \newcommand{\deltap}{0.026983419058523972}
    \newcommand{\pFringe}{0.5}

    \newcommand{\numberxLines}{10}
    \newcommand{\deltax}{0.01}

    \newcommand\USHOCK{1./60}
    \newcommand\SHOCKWIDTH{1./25}
    \newcommand\COMPRESSIONRATIO{4}

    \draw [fill=gray!15, draw=none] (-\shockWidth * \factor, 0) rectangle (\shockWidth * \factor, \gridHeight);

    \foreach \x [evaluate=\x as \xScaled using \x*\factor] in {-\shockWidth, \secondShockStep, ..., \shockWidth}
      {
        \draw (\xScaled,0) -- (\xScaled, \gridHeight*0.4);
        \draw[dotted] (\xScaled, \gridHeight*0.4) -- (\xScaled, \gridHeight*0.6 );
        \draw (\xScaled,\gridHeight*0.6) -- (\xScaled, \gridHeight);x
      }

    \foreach \p [evaluate=\p as \pScaled using \p*\deltap*\factor] in {0, ...,\numberpLines}
      {
        \draw[dotted] (-\gridWidth - \pFringe, \pScaled) -- (-\gridWidth, \pScaled);
        \draw (-\gridWidth, \pScaled) -- (\gridWidth, \pScaled);
        \draw[dotted] (\gridWidth, \pScaled) -- (\gridWidth + \pFringe, \pScaled);

        \draw[dotted] (-\gridWidth - \pFringe, \gridHeight - \pScaled) -- (-\gridWidth, \gridHeight - \pScaled);
        \draw (-\gridWidth, \gridHeight - \pScaled) -- (\gridWidth,\gridHeight - \pScaled);
        \draw[dotted] (\gridWidth, \gridHeight - \pScaled) -- (\gridWidth + \pFringe, \gridHeight -\pScaled);
      }

    \foreach \x [evaluate=\x as \currentx using (sinh(\x * \deltax)+\lastx),
      remember=\currentx as \lastx (initially \shockWidth)] in {1,...,\numberxLines}
      {
        \draw (\currentx * \factor, 0) -- (\currentx *\factor, \gridHeight*0.4);
        \draw[dotted] (\currentx * \factor, \gridHeight*0.4) -- (\currentx *\factor, \gridHeight*0.6);
        \draw (\currentx * \factor, \gridHeight*0.6) -- (\currentx *\factor, \gridHeight);

        \draw (-\currentx * \factor, 0) -- (-\currentx *\factor, \gridHeight*0.4);
        \draw[dotted] (-\currentx * \factor, \gridHeight*0.4) -- (-\currentx *\factor, \gridHeight*0.6);
        \draw (-\currentx * \factor, \gridHeight*0.6) -- (-\currentx *\factor, \gridHeight);
      }

    \draw[-{stealth}] (-\gridWidth, \gridHeight + 0.1) -- (-\gridWidth, \gridHeight + \pFringe) node [anchor=east]{\scriptsize $\ln p^{*}$};

    \node[above] at (1*\stepSizeShock *\factor + 0.5 * \stepSizeShock * \factor, \gridHeight) {\scriptsize $\Delta x^{*}_{\text{s}}$};
    \node[above] at (9.2 * \shockWidth * \factor, \gridHeight) {\scriptsize $\Delta x^{*}_{i}$};
    \node[anchor=east] at (-\gridWidth - \pFringe, \numberpLines * \deltap * \factor - 0.5 * \deltap * \factor ) {\scriptsize $\Delta \ln p^{*} $};

    \begin{axis}[%
      ymajorgrids=true,
      tiny,
      xlabel={$x^{*}$},
      x label style={font=\scriptsize, at={(1.02,0.1)}},
      x = 1cm,
      axis x line=bottom,
      xticklabel={%
          \pgfmathsetmacro{\result}{\tick / \factor}$\pgfmathprintnumber[/pgf/number format/fixed,precision=2]{\result}$},
      axis y line*=right,
      y axis line style={-{stealth},xshift=1cm},
      ylabel={$U^{*}(x^{*})$},
      ytick={1/60, 5.6567e-3, 0.0151766301414,  1/4 * 1/60},
      y tick style={xshift=1cm},
      y tick label style={xshift=1cm, /pgf/number format/precision=4, /pgf/number format/fixed, /pgf/number format/fixed zerofill,},
      scaled y ticks=false,
      y label style={at={(1.1,0.5)}},
      ymin=3.9e-3, ymax=0.0175,
      y grid style={teal!30, xshift=1cm,{shorten <=-1cm} },
      anchor=south, height=\gridHeight cm, scale only axis, ] \addplot[ domain=-\gridWidth:\gridWidth, no markers, samples=250, teal, thick, ] {\USHOCK/(2*\COMPRESSIONRATIO)*((1 - \COMPRESSIONRATIO)*tanh(x/ \factor * \SHOCKWIDTH^(-1)) + (1 + \COMPRESSIONRATIO))};
      \draw[teal!25, thin] (axis cs:-0.5 * \factor,5.6567e-3) -- (axis cs:-0.62*\factor,5.6567e-3);
      \draw[teal!25, thin] (axis cs:-0.5 * \factor,0.0151766301414) -- (axis cs:-0.62*\factor,0.0151766301414);
    \end{axis}

    \draw[ultra thin, gray] (-\shockWidth * \factor,\gridHeight) -- (-\shockWidth * \factor,\gridHeight + 0.09 *\gridHeight) node[above, black] {\tiny $-L^{*}_{s}$};
    \draw[ultra thin, gray] (\shockWidth * \factor, \gridHeight) -- (\shockWidth * \factor,\gridHeight + 0.09 *\gridHeight) node[above, black] {\tiny $L^{*}_{s}$};
  \end{tikzpicture}
  \caption{A detail from the computational grid of the diffusive shock acceleration simulation. It shows the design of the grid around the shock (highlighted in a light grey)  and how it resolves the velocity profile $U(x)$ (drawn in teal).}
  \label{fig:shock-grid}
\end{figure}

Since the change in $p$, i.e. in energy, comes from the derivative in the velocity field $\mathbf{U}$, we cannot use the discontinuous velocity profile~\eqref{eq:discontinuous-velocity-profile}. Instead, we approximate it with
\begin{equation}
  \label{eq:sapphire-velocity-profile}
  U^{*}(x^{*}) = \frac{U^{*}_{1}}{2r}
  \left[1 + r +(1 - r)\tanh(x^{*}/L^{*}_{s})  \right] \,,
\end{equation}
where $L^{*}_{s}$ is the shock width\footnote{For a discussion on the effect of a finite shock-thickness, see \cite{Druryetal82,AchterbergSchure}. The power-law index is modified to $ -\frac{3r}{r-1} - \frac{9}{2(r-1)}\frac{U_1}{V_1} \frac{L_s}{\lambda}$. For the results shown, the correction is $10^{-3}$.}. The shock parameters are chosen such that they plausibly model a supernova remnant shock. A typical speed for such a shock is a few thousand kilometres per second, e.g. $U^{*}_{1}=1/60$. Generally, it is assumed that these shocks are strong, i.e. their compression ratio is $r = 4$. The shock width is chosen to be a fraction of the scattering mean free path, i.e. $L^{*}_{s} = 1/25$. The velocity profile is plotted in Fig.~\ref{fig:shock-grid}.

Since we set the scattering frequency to $\nu^{*} = 1$, the mean free path is $\lambda^{*}= V^{*}/\nu^{*} \approx 1$, i.e. a low energy particle ($\gamma \gtrsim 1$) is scattered once per gyration about the magnetic field and high energy particles are scattered about $T^{*}_{g} \approx \gamma$ times, where $T^{*}_{g}$ is the gyroperiod. In astrophysical plasmas, it is thought that the mean free path of energetic particles will increase with increasing particle energy and, hence, a constant scattering frequency is not a realistic choice. However, it yields the simple form of the exponential term in the analytical solution for the upstream distribution function, see eq.~\eqref{eq:full-distribution-function-upstream}, and reduces simulation times.

As expected on physical grounds, the $\mathbf{B}$-field does not appear in the system of PDEs, see eq.~\eqref{eq:actual-soe-for-sapphire}. Nonetheless, it is included in the simulation and set to $\mathbf{B}^{*}= B^{*}_{0} \mathbf{e}_{x}$, where $B^{*}_{0} = 1$.

The monoenergetic source is modelled as a Gaussian distribution of particles, i.e.
\begin{equation}
  \label{eq:monoenergetic-point-source}
  s(x^{*},p^{*}) = \frac{Q^{*}}{2\pi \sigma^{*}_{p}\sigma^{*}_{x}}
  \exp\left(-\frac{(x^{*} - x^{*}_{0})^{2}}{2 \sigma^{*2}_{x}}\right)\exp\left(-\frac{(p^{*} - p^{*}_0)^{2}}{2 \sigma^{*2}_{p}}\right) \,.
\end{equation}
The spherical harmonic expansion of the source is $s(x^{*},p^{*}) = s_{000}(x^{*},p^{*}) Y_{000}$ with $s_{000} = \sqrt{4\pi} s(x^{*},p^{*})$. Particles are injected directly at the shock, i.e. $x^{*}_{0} = 0$. The rate is $Q^{*} = 0.1$ and the injection momentum is $p^{*}_{0} = 2$. The standard deviations of the Gaussian distribution are $\sigma^{*}_{x} = \sigma^{*}_{p} = 1/8$.

The initial conditions for the expansion coefficients are $\mathbf{f}_{0}(t^{*}=0) = 0$, i.e. initially there are no particles in the computational domain. The boundaries in $x$-direction are treated differently in the up- and downstream region. At the upstream boundary we use the zero inflow boundary condition as described above, and given by eq.~\eqref{eq:definition-boundary-flux}. At the downstream boundary, we expect that the gradient of the asymptotic solution is zero, i.e. $\partial f/\partial x = 0$. We thus allow the inflow to be determined by the values of the approximate solution $\mathbf{f}_{h}$ on the boundary, i.e.
\begin{equation}
  \label{eq:continuous-bc}
  \bm{\mathring{J}}^{B}_{F}(\mathbf{f}_{h})
  = \mat{W}\left(\bm{\Lambda}_{+}\mat{W}^{\transpose}\mathbf{f}_{h}
  + \bm{\Lambda}_{-}\mat{W}^{\transpose}\mathbf{f}_{h}\right) \,.
\end{equation}
We refer to this as the continuous boundary condition. The boundaries in $p$-direction fulfil the zero inflow boundary condition.

The computational domain is $D = [-280, 280] \times [\ln(0.1), \ln(100)]$. We require the spatial grid to cover multiple diffusion lengths, i.e. $L^{*}_{d} = \lambda^* V^{*}_{1}/(3 U^{*}_{1}) \approx 20 \approx x^{*}_{max} / 14$. The dimension in the $p$-direction covers multiple orders of magnitude to show that \sapphire produces an extended power law.

To resolve the shock region accurately, we adapted the cell size in $x$-direction, see Fig.~\ref{fig:shock-grid}. In the shock region (highlighted in grey) we chose a constant cell size $\Delta x^{*}_{\text{s}} = 0.01$ and outside it the cell size increases as $ \Delta x^{*}_{i} = \sinh(i * 0.01)$. The coarse resolution in the outer parts of the domain allows us to simulate large upstream and downstream regions. The cell size in $\ln p$-direction is $\Delta \ln p^{*} = (\ln (100) - \ln(0.21))/256 \approx 0.027$. For the time evolution we use the implicit Crank--Nicolson method with time step $\Delta t^{*} = 1$. The simulation is run up to a final time of $t^{*}_{F} = 5 \times 10^{5}$ to achieve steady--state.\footnote{We confirmed the results using an explicit fourth order Runge--Kutta (ERK4) method. But as it requires a much smaller time step, we terminated the simulation at an earlier time.}

An overview of the simulation parameters is collected in Tab.~\ref{tab:sapphire-simulation-parameters}.

\begin{table}
  \centering
  \caption{Simulation parameters modelling a supernova remnant shock in \sapphire.}
  \label{tab:sapphire-simulation-parameters}
  \begin{tabular}{lll}
    \hline
    Parameter        & Value                                    & Description                                                         \\
    \hline
    $U^{*}_{1}$      & 1/60                                     & The velocity of the supernova remnant shock                         \\
    $r$              & 4                                        & The compression ratio of the shock                                  \\
    $L^{*}_{s}$      & 1/25                                     & The width of the shock's velocity profile                           \\
    $\nu^{*}$        & 1                                        & \makecell[l]{The scattering frequency describing the rate of        \\ particle-wave interactions}         \\
    $B^{*}_{0}$      & 1                                        & Strength of the magnetic field                                      \\
    $Q^{*}$          & 0.1                                      & \makecell[l]{The injection rate at the shock in number of particles \\ per unit dimensionless time} \\
    $p^{*}_{0}$      & 2                                        & The injection momentum of the particles                             \\
    $x^{*}_{0}$      & 0                                        & Location of the injection                                           \\
    $\sigma^{*}_{x}$ & 1/8                                      & The width of the source in $x$-direction                            \\
    $\sigma^{*}_{p}$ & 1/8                                      & The width of the source in $p$-direction                            \\
    $D$              & $[-280,280]  \times[\ln(0.1), \ln(100)]$ & The computational domain                                            \\
    $\Delta t^{*}$   & 1                                        & The time step size                                                  \\
    $ t^{*}_{F}$     & $5 \times 10^{5}$                        & The final time of the simulation                                    \\
    \hline
  \end{tabular}
\end{table}

\paragraph{Results}
\label{sec:simulation-results}

\begin{figure}
  \centering

  \begin{tikzpicture}[baseline]
    \begin{loglogaxis}[
        axis lines=left,
        title={$x^{*} = \text{const.} = 0$},
        xlabel={$p^{*} = p/\underline{p}$},
        xmin=0.8, xmax=100,
        ymin=1e-6,
        grid = both,
        width=\textwidth*0.41,
        legend entries={$f(p^{*})$},
        legend style={at={(0,1.03)}, anchor=south east, draw=none},
        cycle list name=exotic,
      ]
      \newcommand\USHOCK{1./60}
      \newcommand\COMPRESSIONRATIO{4}
      \newcommand\INJECTIONMOMENTUM{2.}
      \newcommand\SOURCE{1./10}
      \newcommand\SIGMAP{1./8}
      \newcommand\SIGMAX{1./8}

      \addplot+ [solid, no markers, semithick] table[x index=4, y index=0, col sep=comma] {fp.csv};

      \addplot [domain=0.1:100, dashed]{\SOURCE/(\USHOCK * \INJECTIONMOMENTUM) *  (3 * \COMPRESSIONRATIO)/(\COMPRESSIONRATIO - 1)
        * (x / \INJECTIONMOMENTUM )^(-4)};

      \addplot [gray, dotted, thick, domain=0.1:4, samples=200] {1/(2*pi * \SIGMAX * \SIGMAP) * exp(-(x - \INJECTIONMOMENTUM)^2/(2 * \SIGMAP^2))};
      \node[gray] at (4.4,1e-4) {$s(0, p^{*})$};
    \end{loglogaxis}
  \end{tikzpicture}

  \hskip 5pt

  \begin{tikzpicture}[baseline]
    \newcommand\USHOCK{1./60}
    \newcommand\SHOCKWIDTH{1./25}
    \newcommand\COMPRESSIONRATIO{4}
    \newcommand\SCATTERINGFREQ{1.}
    \newcommand\MOMENTUM{10.}
    \newcommand\VELOCITY{1/sqrt(1 + \MOMENTUM^(-2))}
    \newcommand\INJECTIONMOMENTUM{2.}
    \newcommand\SOURCE{1./10}

    \begin{semilogyaxis}[
        axis y line=left,
        axis x line=bottom,
        title ={$p^{*} = \text{const.} = 10$},
        xlabel = {$x ^{*} = x/\underline{r}_{g}$},
        xmin=-280, xmax=280,
        legend style={at={(0,1.03)}, anchor=south east, draw=none},
        scaled y ticks = false,
        xmajorgrids,
        ymax= 2e-1,
        minor y tick num = 0,
        width=\textwidth*0.41,
        cycle list name=exotic,]

      \addplot+ [solid, no markers, semithick] table[x index=2, y index=0, col sep=comma,] {fx.csv};
      \addlegendentry{$f(x^{*})$}

      \addplot [%
        forget plot,
        black,
        dashed,
        domain=0:279,
        samples=100,
      ] { \SOURCE/(\USHOCK * \INJECTIONMOMENTUM) *  (3 * \COMPRESSIONRATIO)/(\COMPRESSIONRATIO - 1)
        * (\MOMENTUM / \INJECTIONMOMENTUM )^(-4)};

      \addplot [%
        forget plot,
        black,
        dashed,
        domain=-279:0,
        samples=200,
      ] {\SOURCE/(\USHOCK * \INJECTIONMOMENTUM)
        *  (3 * \COMPRESSIONRATIO)/(\COMPRESSIONRATIO - 1) * (\MOMENTUM / \INJECTIONMOMENTUM )^(-4)
        * exp((3*\SCATTERINGFREQ * \USHOCK * x)/ \VELOCITY^2)};
    \end{semilogyaxis}

    \begin{axis}[%
        axis y line=right,
        axis x line=none,
        yticklabel style={/pgf/number format/precision=4, /pgf/number format/fixed, /pgf/number
            format/fixed zerofill,},
        scaled y ticks=false,
        legend style={at={(1,1.03)}, anchor=south west, draw=none, legend plot pos=right},
        ymin=-0.0010, ymax=0.0002,
        xmin=-280, xmax=280,
        width=\textwidth*0.41,
      ]

      \addplot+ [solid, no markers, semithick, orange] table[x index=2, y index=4, col sep=comma] {fx.csv};
      \addlegendentry{$a(x^{*})$}

      \addplot [%
        forget plot,
        black,
        dashed,
        domain=0:279,
        samples=100,
      ] {0};

      \addplot[
        forget plot,
        black,
        dashed,
        domain=-279:0,
        samples=100,
      ] {- (3 * \USHOCK) * \VELOCITY^(-1) * \SOURCE/(\USHOCK * \INJECTIONMOMENTUM)
        *  (3 * \COMPRESSIONRATIO)/(\COMPRESSIONRATIO - 1) * (\MOMENTUM / \INJECTIONMOMENTUM )^(-4)
        * exp((3*\SCATTERINGFREQ * \USHOCK * x)/ \VELOCITY^2)};

      \addplot[
        gray,
        dotted,
        thick,
        domain=-279:279,,
        samples=250,
      ] {\USHOCK/(2*\COMPRESSIONRATIO)*((1 - \COMPRESSIONRATIO)*tanh(x * \SHOCKWIDTH^(-1)) + (1 + \COMPRESSIONRATIO))};
      \node [gray, anchor=south east] at (200,0.005) {$U^{*}(x^{*})$};
    \end{axis}

    \begin{axis}[%
        gray,
        dotted,
        axis y line=none,
        axis x line=none,
        scaled y ticks=false, ytick={1/60, 1/4 * 1/60}, yticklabels={$1/60$, $1/240$}, ymin=0, ymax=0.02, xmin=-280, xmax=280, width=\textwidth*0.41, ] \addplot[ gray, dotted, thick, domain=-279:279,, samples=250, ] {\USHOCK/(2*\COMPRESSIONRATIO)*((1 - \COMPRESSIONRATIO)*tanh(x * \SHOCKWIDTH^(-1)) + (1 + \COMPRESSIONRATIO))}; \node [gray, anchor=south east] at (200,0.005) {$U(x^{*})$};
    \end{axis}
  \end{tikzpicture}
  \caption{Comparison between analytical solution and numerical solution. Left panel: A log-log plot of the particle spectrum at the shock. Right panel: A plot of the isotropic and anisotropic part of the distribution function for a constant momentum. \\
  Coloured plots present the numerical results and the dashed plots show the analytical solution. \\ The dotted plots show the width of the source term $s(0,p^{*})$ (left panel) and the velocity profile $U^{*}(x^{*})$ of the shock in its rest frame (right panel).\\
  The units of the distribution function are number of particles per unit dimensionless length and dimensionless momentum. }
  \label{fig:comparison-analytic-numerical-solution}
\end{figure}
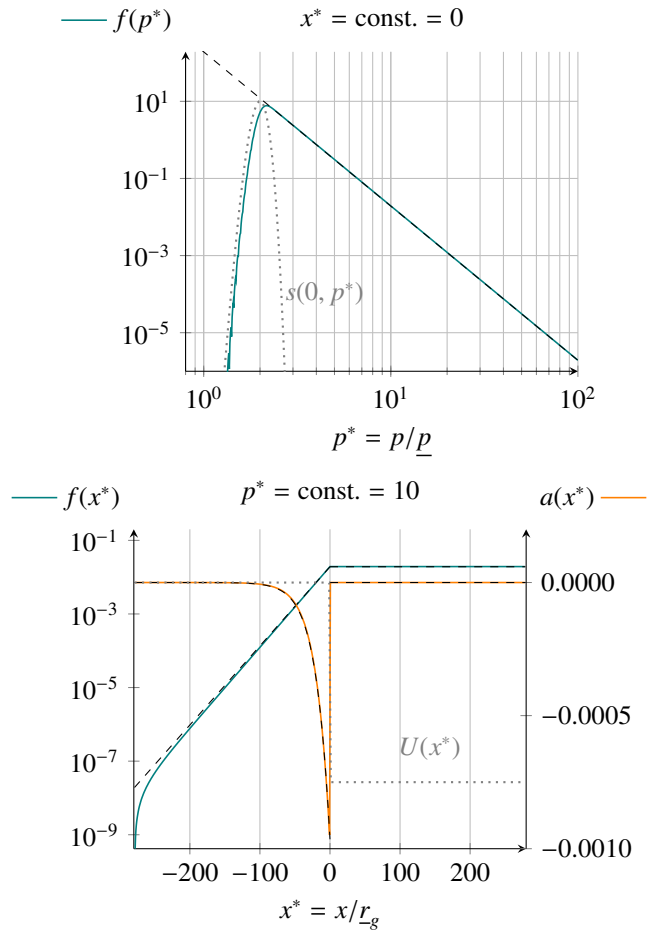

In Fig.~\ref{fig:comparison-analytic-numerical-solution} we compare the steady-state analytic solution with the numerical solution computed with \sapphiren. In the left panel, the numerically computed spectrum at the shock is compared to the analytic expectation given in eq.~\eqref{eq:particle-spectrum-shock-rest-frame}, i.e. a power law with spectral index $\alpha = -3r/(r - 1) = -4$ and normalisation $N \coloneqq Q^{*}/(p^{*}_{0} U^{*}_{1}) 3r/(r - 1) = 12$. The spectral index of the numerical solution is $|\alpha_{\text{num}}-4| = 1.5\times 10^{-3}$, and it extends up to $p*=100$, which is the boundary of the computational domain in the $p$-direction. The resolution of the ordinate of the log-log plot is too low to see that the normalisation of the numerical solution $N_{\text{num}} = 12.14$ is off by $(N - N_{\text{num}})/N \approx 0.012 = 1.2 \%$. We speculate that this discrepancy is due to approximating a point injection of particles with a Gaussian distribution.

In the right panel, we compare the spatial profile of the numerical solutions with the analytic solution given in eq.~\eqref{eq:full-distribution-function-upstream} and \eqref{eq:full-distribution-function-downstream} and evaluated at $p^{*}= 10$. The discrepancy between the computed isotropic part and the analytical result at the left boundary of the spatial domain is due to the boundary condition that enforces zero inflow. There is also a small difference in the downstream normalisation, which is not visible due to the log scaling of the $f(x^{*})$-axis. This is the same discrepancy as in the normalisation of the particle spectrum discussed in the previous paragraph.

\begin{figure}
  \centering
  \includegraphics{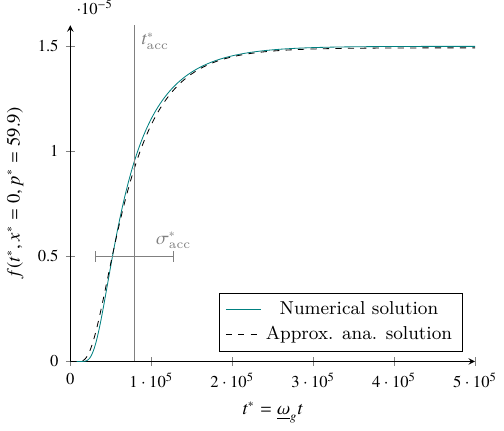}
  \caption{The plot shows the temporal evolution of the isotropic part of the particle spectrum $f$ at $x^*=0$ and $p^{*} = 59.9$. The orange curve shows the approximate analytic expression and the teal-coloured curve shows the numerical result. \\
  The mean acceleration time is $t^{*}_{\text{acc}} = 78909.74$ and the standard deviation is $\sigma^{*}_{\text{acc}} = 48859.1$.
  }
  \label{fig:temporal-evolution-spectral-index}
\end{figure}

In Fig.~\eqref{fig:temporal-evolution-spectral-index} we plot the temporal evolution of the numerically computed spectrum at the shock's position for a fixed momentum $p^{*} = 59.9$ and compare it with the \emph{approximate} analytic expression given in the eqs.~\eqref{eq:approximate-temporal-evolution-spectrum}--\eqref{eq:cumulative-temporal-distribution}. Despite the fact that the setup used in \sapphire only approximates the assumptions leading to the steady-state spectrum and its time-dependent counterpart and the fact that the analytic expression for the temporal evolution is also merely an approximation, the two curves follow each other closely. This indicates that the temporal evolution of the spectrum is captured accurately by \sapphire.

\section{Conclusions}
\label{sec:conclusions}

We introduced a new VFP solver called \sapphire whose distinguishing feature is the combination of a spherical harmonic expansion of the distribution function \emph{and} the application of the discontinuous Galerkin method to compute the expansion coefficients numerically. The motivation is to exploit knowledge about the distribution function in specific astrophysical environments and about the VFP equation; namely the fact that the distribution of particles is almost isotropic in some environments and that the VFP equation is an advection-reaction equation. Moreover, solving a kinetic equation like the VFP equation avoids difficulties such as statistical noise, which need to be addressed in discrete sampling approaches like the PIC method \cite[e.g.][]{BirdsallLangdon}.

The spherical harmonic expansion of the distribution function leads to a system of PDEs given in eq.~\eqref{eq:real-system-of-equations}. The system can be formulated in a convenient way using representation matrices of operators, which is derived and described in \cite{Schween24a}. The relevant point of the new formulation is that it brings out the advection-reaction character of the PDE system, which we made explicit in eq.~\eqref{eq:system-as-advection-reaction-eq}. Advection-reaction equation are particularly amenable to the application of the discontinuous Galerkin method.

The discontinuous Galerkin method is the key algorithm underlying \sapphire. We explained in detail how to apply it to the system of PDEs. A highlight is the upwind flux, i.e. the properties of the advection matrices $\mat{\beta}^{a}$ allowed us to use an upwind flux without computing the eigenvectors and eigenvalues at each cell interface.

To validate \sapphire we presented a number of examples where exact or stationary solutions are known. It was demonstrated that the code can be run with explicit or implicit time stepping, and satisfies the expected convergence scaling, both temporally and spatially. The difference between the physical solution and the mathematical solution with finite expansion was emphasised. The quality of the solution with increasing expansion order of spherical harmonics was explored for a simple test case in the collisionless limit.

Finally, we simulated the acceleration of particles at a parallel shock, comparing the numerical results with the well-known test-particle solution \cite[e.g.][]{Drury1983,Drury1991} \cite[see also][]{BlandfordEichler, Jonesellison, Kirk_ParticleAcc}. For comparison with analytic solutions in the literature, we adopted a uniform momentum independent scattering rate. \sapphire can simulate a momentum dependent scattering rate, though reducing the rate at higher energies increases rapidly the acceleration time, and hence also the computational cost.

The simulations were performed with expansion order $l_{\text{max}} = 1$, which is acceptable for strictly parallel, non-relativistic shocks. If the angle between the shock normal and the mean field are misaligned, higher-order terms are necessary (see for example the discussion in \cite{Bell2011} and \cite{TakamotoKirk}). Since \sapphire is not restricted to $l_{\text{max}} = 1$, it can be used to simulate particle acceleration at oblique shocks.

We conclude that the spherical harmonic expansion method to solve the Vlasov-Fokker-Planck equation is well-suited to the discontinuous Galerkin approach. In future versions, \sapphire will implement adaptive mesh refinement capabilities provided by the \texttt{deal.ii} library. Because of its ``locality'' the dG method is ideal for these kinds of algorithms. From an astrophysics perspective, the most interesting extension of \sapphire is the inclusion of a ``fluid module'', i.e. a self-consistent computation of the velocity $\mathbf{U}$ and the magnetic field $\mathbf{B}$ of the background plasma. Such a module will open countless possibilities to study the self-consistent feedback of energetic particles onto the background plasma \cite{Reville2013, Belletal13}.

\sapphire has been developed with applications relevant to the high-energy astrophysics community in mind. As a free and open-source software, the range of possible applications can be broadened. In its current form, a limitation of its applicability concerns the choice of scattering operator. Extensions of the code for modelling of laboratory plasmas would require a more sophisticated scattering operator, such as the that implemented in codes used in inertial confinement fusion studies \cite[e.g.][]{Tzoufras2011,Thomas2012,Bell24}.

\section*{Declaration of competing interest}
The authors declare that they have no known competing financial interests or personal relationships that could have appeared to influence the work reported in this paper.

\section*{Data availability}
No data was used for the research described in the article.
\sapphire can be downloaded from the git repository  {\href{https://github.com/sapphirepp/sapphirepp}{\faGithub}}\,.
All results shown can be reproduced with the examples provided therein.

\section*{Acknowledgements}
The authors would like to thank Dr. Philipp Gerstner who was kind and patient enough to introduce us to the world of finite elements and the dG method in particular. Furthermore, we are very grateful to Prof. John Kirk for his knowledgeable feedback.

\appendix
\section{Definition of the real spherical harmonics}
\label{app:definition-real-spherical-harmonics}

\sapphire solves the system of equations~\eqref{eq:real-system-of-equations} and, thus, it computes the expansion coefficients $f_{lms}$. For a physical interpretation of the results, a reconstruction of the distribution function $f$ may be useful. This requires to know how the real spherical harmonics are defined. They are defined to be
\begin{equation}
  \label{eq:spherical-harmonics}
  Y_{lms}(\theta, \varphi) \coloneqq N_{lm} P^{m}_{l}(\cos\theta)
  \left(\delta_{s0}\cos{m\varphi} + \delta_{s1}\sin{m\varphi}\right) \,.
\end{equation}
Where $N_{lm}$ is a normalisation, which is
\begin{equation}
  \label{eq:normalisation-spherical-harmonics}
  N_{lm} = \sqrt{\frac{2l + 1}{2\pi(1 + \delta_{m0})} \frac{(l-m)!}{(l+m)!}} \,,
\end{equation}
and the functions $P^{m}_{l}$ are the \textit{associated Legendre Polynomials}. Their definition is given in ~\cite[eq. 8.6.6]{Stegun_HandbookOfFunctions} as
\begin{equation}
  \label{eq:associated-legendre-polynomials}
  P^{m}_{l}(\cos\theta) \coloneqq
  (-1)^{m}\sin^{m}\theta \frac{\mathrm{d}^{m}}{\mathrm{d}(\cos\theta)^{m}}P_{l}(\cos\theta) \,.
\end{equation}
Note that the \textit{Condon--Shortley} phase $(-1)^{m}$ is included in the definition of the associated Legendre polynomials and \emph{not} in the definition of the spherical harmonics. $P_{l}$ is the \textit{Legendre polynomial} of degree $l$. A definition of the Legendre polynomial $P_{l}$ is given through
\begin{equation}
  \label{eq:legendre-polynomials}
  P_{l}(\cos\theta) \coloneqq \frac{1}{2^{l}l!} \frac{\mathrm{d}^{l}}{\mathrm{d}(\cos\theta)^{l}}(\cos^{2}\theta - 1)^{l} \,,
\end{equation}
which can, for example, be found in~\cite[eq. 8.6.18]{Stegun_HandbookOfFunctions}.

The real spherical harmonics relate to the complex spherical harmonics
\begin{equation}
  Y^{m}_{l}(\theta, \varphi) = N^{m}_{l} P^{m}_{l}(\cos\theta)e^{\complexi m \varphi}
  \quad \text{with }   N^{m}_{l} = \sqrt{\frac{2l + 1}{4\pi} \frac{(l-m)!}{(l+m)!}}
\end{equation}
through
\begin{align}
  Y_{l, m=0, s=0}(\theta, \varphi) & = Y^{m=0}_{l}(\theta, \varphi)                                                                      &  & \text{for } m = 0, s=0   \\
  Y_{l, m, s=0}(\theta, \varphi)   & = \frac{1}{\sqrt{2}}\left(Y^{m}_{l}(\theta, \varphi) + (-1)^{m} Y^{-m}_{l}(\theta, \varphi)\right)  &  & \text{for } m \neq 0     \\
  Y_{l, m, s=1}(\theta, \varphi)   & = \frac{1}{\sqrt{2}i}\left(Y^{m}_{l}(\theta, \varphi) - (-1)^{m} Y^{-m}_{l}(\theta, \varphi)\right) &  & \text{for } m \neq 0 \,,
\end{align}
or in short,
\begin{equation}
  Y_{lms}(\theta, \varphi) = \frac{1}{\sqrt{2 (1 + \delta_{m0})}} (-i)^{s} \left(Y^{m}_{l}(\theta, \varphi) + (-1)^{s} {Y^{m}_{l}}^{*}(\theta, \varphi)\right) \,.
\end{equation}

And we have the following relation,
\begin{equation}
  \int_{S^{2}} Y_{l'm's'} Y_{lms} \mathrm{d}\Omega = \delta_{l'l}\delta_{m'm}\delta_{s's} \,.
\end{equation}

We give explicit expressions for the first few real spherical harmonics in Tab.~\ref{tab:real-spherical-harmonics}.

\begin{table}
  \centering
  \caption{List of real spherical harmonics $Y_{lms}(\theta, \varphi)$ for $l \leq 2$.}
  \label{tab:real-spherical-harmonics}
  \begin{tabular}{ll}
    \hline
    $Y_{000}(\theta, \varphi)$ & $\sqrt{\frac{1}{4 \pi}}$                                              \\
    $Y_{100}(\theta, \varphi)$ & $\sqrt{\frac{3}{4 \pi}} \cos\theta$                                   \\
    $Y_{110}(\theta, \varphi)$ & $-\sqrt{\frac{3}{4 \pi}} \sin\theta \cos\varphi$                      \\
    $Y_{111}(\theta, \varphi)$ & $-\sqrt{\frac{3}{4 \pi}} \sin\theta \sin\varphi$                      \\
    $Y_{200}(\theta, \varphi)$ & $\frac{1}{4}\sqrt{\frac{5}{\pi}} \left(3 \cos^2\theta - 1\right)$     \\
    $Y_{210}(\theta, \varphi)$ & $\frac{-1}{2}\sqrt{\frac{15}{\pi}} \sin\theta \cos\theta \cos\varphi$ \\
    $Y_{211}(\theta, \varphi)$ & $\frac{-1}{2}\sqrt{\frac{15}{\pi}} \sin\theta \cos\theta \sin\varphi$ \\
    $Y_{220}(\theta, \varphi)$ & $\frac{1}{4}\sqrt{\frac{15}{\pi}} \sin^2\theta \cos{2\varphi}$        \\
    $Y_{221}(\theta, \varphi)$ & $\frac{1}{4}\sqrt{\frac{15}{\pi}} \sin^2\theta \sin{2\varphi}$        \\
    \hline
  \end{tabular}
\end{table}

\section{Higher order corrections}
\label{app:high-order}

As mentioned, dropping the relativistic corrections in front of the time derivative is accurate to order $(U/V)$. In this appendix we want to demonstrate two different ways, to retain higher order corrections in $(U/V)$ in \texttt{Sapphire++}.

Starting from the VFP equation in mixed coordinates \eqref{eq:vfp-mixed-coordinates},
\begin{equation}
  \label{eq:vfp-mixed-coordinates-appendix}
  \left(1+\frac{\mathbf{U}\cdot \mathbf{V}'}{c^2}\right)\frac{\partial f}{\partial t}
  + \left(\mathbf{U} + \mathbf{V}'\right) \cdot \nabla_{x} f
  - \left(\gamma' m \frac{\mathrm{d} \mathbf{U}}{\mathrm{d} t}
  + (\mathbf{p}' \cdot \nabla_{x}) \mathbf{U} \right) \cdot \nabla_{p'} f
  + q \mathbf{V}' \cdot \left(\mathbf{B}' \times \nabla_{p'} f \right)
  = \frac{\nu'}{2}\Delta_{\theta', \varphi'}f \,,
\end{equation}
we apply the same operator based method \cite{Schween24a} to arrive at the following system,
\begin{equation}
  \begin{split}
    \left(\mat{1} + \frac{V}{c^2} U_{a} \mat{A}^{a}\right) \partial_{t}\mathbf{f}
     & + \left(U^{a}\mat{1} + V \mat{A}^{a}\right) \partial_{x^{a}}\mathbf{f}
    - \left(\gamma m \frac{\mathrm{d} U_{a}}{\mathrm{d} t} \mat{A}^{a}
    + p \frac{\partial U_{b}}{\partial x^{a}} \mat{A}^{a}\mat{A}^{b}\right)\partial_{p} \mathbf{f}             \\
     & {} + \left(\frac{1}{V} \epsilon_{abc} \frac{\mathrm{d} U^{a}}{\mathrm{d} t} \mat{A}^{b}\mat{\Omega}^{c}
    + \epsilon_{bcd} \frac{\partial U_{b}}{\partial x^{a}}\mat{A}^{a}\mat{A}^{c}\mat{\Omega}^{d}\right) \mathbf{f}
    - \omega_{a}\mat{\Omega}^{a}\mathbf{f}
    + \nu \mat{C}\mat{f} = 0 \,.
  \end{split}
\end{equation}
Using dG and an explicit Euler step to discrete the equations (compare equation \eqref{eq:theta-method}),
\begin{equation}
  \tilde{\mat{M}} \frac{\bm{\zeta}^{n} - \bm{\zeta}^{n-1}}{\Delta t} =
  \mathbf{h}^{n-1} - \mat{D}^{n-1} \bm{\zeta}^{n-1} \, ,
\end{equation}
we introduce the modified mass matrix
\begin{equation}
  (\tilde{\mat{M}})_{ij}   \coloneqq \sum_{T \in \mathcal{T}_{h}} \int_{T} \bm{\phi}_{i} \left(\mat{1} + \frac{V}{c^2} U_{a} \mat{A}^{a}\right) \bm{\phi}_{j} \, .
\end{equation}
Solving the system of PDEs involves solving a linear system of equations given by the modified mass matrix. This computationally more expensive than solving the system of equations corresponding to the ordinary mass matrix, because the modified mass matrix is less sparse. This statement holds true for explicit time stepping. For implicit time steps, the linear system of equations is more complex and incorporating the modified mass matrix should not affect the solver.

In a different approach, we can multiply \eqref{eq:vfp-mixed-coordinates-appendix} through by $ 1-\frac{\mathbf{U}\cdot \mathbf{V}'}{c^2}$, and neglect terms of order $\mathcal{O}\left((U/V)^2\right)$,
\begin{equation}
  \frac{\partial f}{\partial t}
  + \left(\mathbf{U} + \mathbf{V}' - \mathbf{V}' \frac{\mathbf{U}\cdot \mathbf{V}'}{c^2}\right) \cdot \nabla_{x} f
  - \left(\gamma' m \frac{\mathrm{d} \mathbf{U}}{\mathrm{d} t}
  + (\mathbf{p}' \cdot \nabla_{x}) \mathbf{U} \right) \cdot \nabla_{p'} f
  + q \left(1-\frac{\mathbf{U}\cdot \mathbf{V}'}{c^2}\right) \mathbf{V}' \cdot \left(\mathbf{B}' \times \nabla_{p'} f \right)
  = \frac{\nu'}{2} \left(1-\frac{\mathbf{U}\cdot \mathbf{V}'}{c^2}\right) \Delta_{\theta', \varphi'}f \,.
\end{equation}
Applying the operator based method, we arrive at the following system,
\begin{equation}
  \begin{split}
    \partial_{t}\mathbf{f}
     & + \left(U^{a}\mat{1} + V \mat{A}^{a} - \frac{V^2}{c^2} U_{b} \mat{A}^{a}\mat{A}^{b}\right) \partial_{x^{a}}\mathbf{f}
    - \left(\gamma m \frac{\mathrm{d} U_{a}}{\mathrm{d} t} \mat{A}^{a}
    + p \frac{\partial U_{b}}{\partial x^{a}} \mat{A}^{a}\mat{A}^{b}\right)\partial_{p} \mathbf{f}                           \\
     & {} + \left(\frac{1}{V} \epsilon_{abc} \frac{\mathrm{d} U^{a}}{\mathrm{d} t} \mat{A}^{b}\mat{\Omega}^{c}
    + \epsilon_{bcd} \frac{\partial U_{b}}{\partial x^{a}}\mat{A}^{a}\mat{A}^{c}\mat{\Omega}^{d}\right) \mathbf{f}
    - \left(\mat{1} - \frac{V}{c^2} U_{b}\mat{A}^{b}\right) \omega_{a}\mat{\Omega}^{a}\mathbf{f}
    + \nu \left(\mat{1} - \frac{V}{c^2} U_{a}\mat{A}^{a}\right) \mat{C}\mat{f} = 0 \,.
  \end{split}
\end{equation}
Computing the upwind flux for the term $\propto U_{b}\mat{A}^{a}\mat{A}^{b}\partial_{x^{a}}$ requires the eigenvalues and eigenvectors of the combined matrix $U_{b}\mat{A}^{a}\mat{A}^{b}$. So far, we are not aware of an analytical solution for this. A numerical solution (similar to the term $\propto \frac{\partial U_{b}}{\partial x^{a}} \mat{A}^{a}\mat{A}^{b}\partial_{p}$) is again computationally expensive.

Finally, we note that another route is to perform the computation in the laboratory frame, investigating other forms of the scattering operator.

\bibliographystyle{elsarticle-num}
\bibliography{references.bib}

\end{document}